\documentclass[fleqn,usenatbib]{mnras}

\usepackage{newtxtext,newtxmath}
\usepackage{lscape} 
\usepackage[T1]{fontenc}
\usepackage{longtable}
\DeclareRobustCommand{\VAN}[3]{#2}
\let\VANthebibliography\thebibliography
\def\thebibliography{\DeclareRobustCommand{\VAN}[3]{##3}\VANthebibliography}

\usepackage{graphicx}	
\usepackage{amsmath}	
\usepackage{subcaption}

\title[RCFM]{  Rotation Curve Fitting Model}

\author[S. N. Cisneros et al.]{
Sophia Natalia Cisneros,$^{1}$\thanks{E-mail: sofcis94@uw.edu}
Richard Ott,$^{2}$
Meagan Crowley,$^{3}$
Amy Roberts,$^{4}$ 
Marcus Paz,$^{5}$
Zaneeyiah Brown,$^{5}$\\
\newauthor
Phong Pham,$^{5}$
Zac Holland,$^{5}$
Robert Real Rico,$^{5}$
Elizabeth Gutierrez-Gutierrez,$^{5}$
Landon Joyal,$^{5}$\\
\newauthor
Amanda Livingston,$^{5}$
Lily Castrellon,$^{5}$
Summer Graham,$^{5}$
Shanon J. Rubin,$^{6}$
Aaron Ashley,$^{2}$
Dillon Battaglia,$^{2}$\\
\\
\newauthor
Daniel Lopez,$^{1}$
and Maya Salwa$^{1}$\\
$^{1}$Department of Astronomy, University of Washington, 1410 NE Campus Pkwy, Seattle, 98195, USA\\
$^{2}$Department of Physics, Massachusetts Institute of Technology, 77 Massachusetts Ave, Cambridge,  02139, USA\\
$^{3}$Department of Physics, University of Massachusetts Boston, 100 Morrissey Boulevard, Boston,  02125, USA\\
$^{4}$Department of Physics, University of Colorado Denver, 1201 Larimer St, Denver,  80204, USA\\
$^{5}$Department of Physics, University of Denver, 2199 S University Blvd, Denver, 80210, USA\\
$^{6}$Department of Physics, Boston University,  233 Bay State Road, Boston,  02215, USA
}

\date{Accepted XXX. Received YYY; in original form ZZZ}

\pubyear{2024}

\begin{document}
\label{firstpage}
\pagerange{\pageref{firstpage}--\pageref{lastpage}}
\maketitle

\begin{abstract}
One key piece of evidence for dark matter is the flat rotation curve problem: the disagreement between measured galactic rotation curves and their luminous mass. 
A novel solution to this problem is presented here. 
A model of relativistic frame effects on Doppler shifts due to the slightly curved frames of emitting galaxies with respect to the Milky Way frame is derived. 
This model predicts   observed Doppler shifted spectra  based only on the   luminous matter estimates and one free model parameter. 
The rotation curve fitting model presented  is tested 
  on  the SPARC sample  of 175   galactic rotation profiles, also  fitted by
  also fitted by dark matter models, and MOND via the Radial Acceleration Relation.  For the SPARC sample, 
  the  new rotation curve fitting model  gives an 
  averaged reduced $\chi^2_r = 2.25$ for 172 galaxies fitted,  
  the isothermal dark matter model gives $\chi^2_r = 1.90$   for   165 galaxies fitted, 
  and  
  the Radial Acceleration Relation (MOND precursor) gives    $\chi^2_r = 4.22$ for 175 galaxies fitted. The model presented here has a free parameter that is highly correlated with a ratio of photometric quantities. 
Implications of this model are discussed. 
\end{abstract}




\section{Introduction} 
\label{sec:intro}
 
\begin{figure}
    \includegraphics[width=\columnwidth]{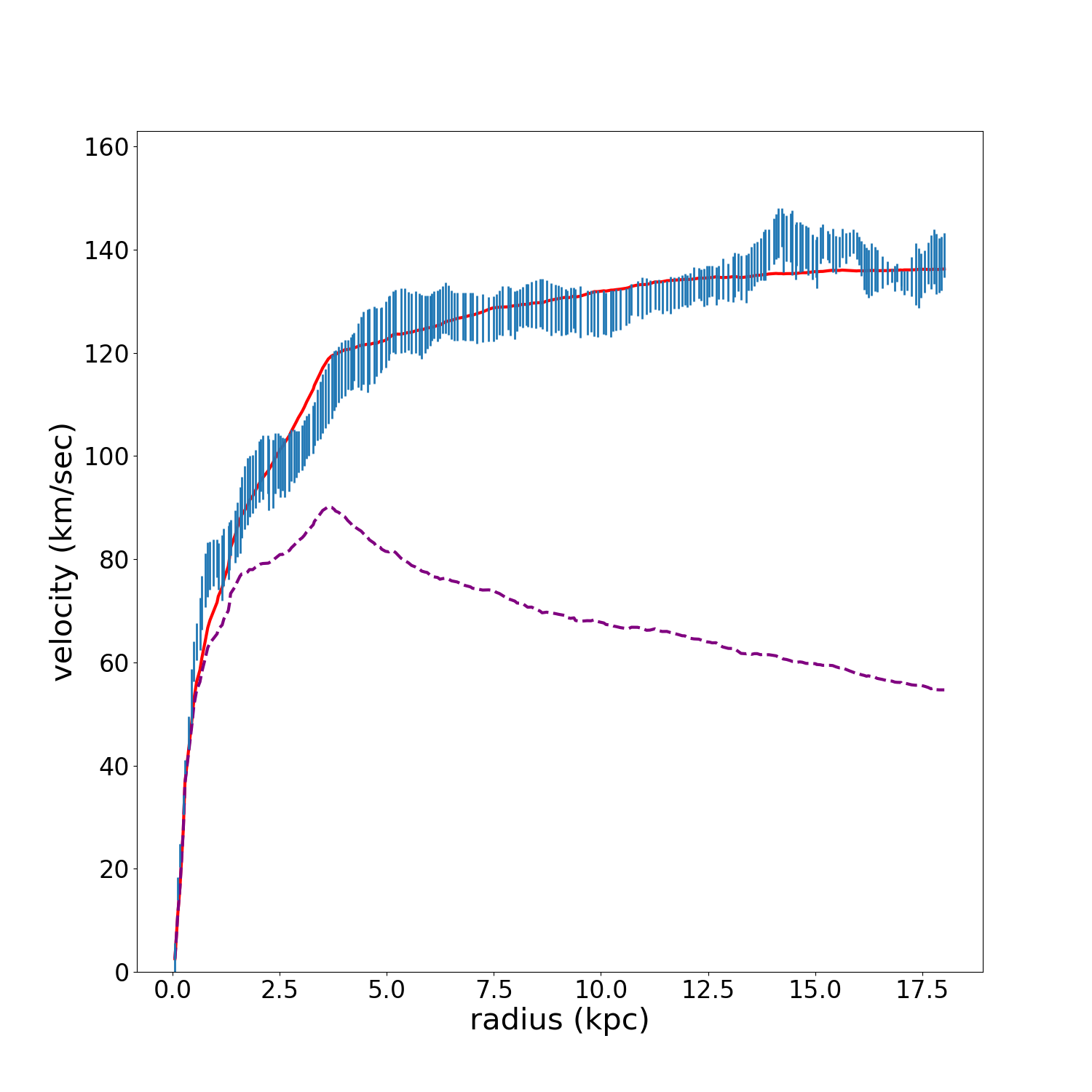}
    \caption{ NGC 2403.  Blue points with error bars: Rotation curve data.  Red line:  RCFM fit.  Dashed purple line: luminous mass model. Rotation curve data and luminous mass models from  de Blok et al. (2008a) for NGC 2403 and  Sofue-Xue-Jiao for the    Milky Way.} 
\label{fig:NGC2403RCFM}
\end{figure}

The flat-rotation curve problem  is  the divergence of two  rotation velocities about the center of a spiral galaxy,  inferred from   different observations of light \citep{Rub,Bosma,1985ApJAlbada}: photometry and Doppler shifted spectra.  
Photometry gives estimates of the luminous mass, which when interpreted classically by the Poisson equation gives the   expected Keplerian orbital  velocities which decline  beyond    the stars. 
 Doppler shifted   spectra, however,  give the 
``flat-rotation curve'' velocities, which   remain  essentially constant far past the stars. 
The divergence of these two velocities  has been
 primary evidence for dark matter theories (Fig.~\ref{fig:NGC2403RCFM}). 
 
 Dark matter  is   hypothesized to be    massive particles which are electromagnetically neutral, and    have a very low interaction probability with baryonic matter, but otherwise obey classical gravity. Though  direct detection experiments continue, in the absence of  a definitive observation of such particles \citep{Cebrian:2022brv},   the phenomenological details  of dark matter problems remain    interesting.   
 S. McGaugh  notes a degeneracy between the luminous and dark matter components in a galaxy,  when he asks    ``Why is the luminous tail wagging the dark matter dog,  if dark matter dominates dynamics?'' \citep{1999McGaugh}. This references the   curious fact   that knowledge  of  the   stellar   disk   completely  determines the spherical dark matter halo, even though dark matter is postulated to dominate dynamics \citep{2004ApJ...609..652M}.

Another interesting  trend is   the Universal Rotation Curve (URC), in which  
a spectrum of   1,100  rotation curves (RCs) inflect about the presumed rotation curve of the Milky Way  \citep{1978Rubin, salucci,  10.1111/j.1365-2966.2007.11696.x}.
In the URC spectrum, galaxy RCs
 are separated   with respect to the luminous mass of the Milky Way, where the RCs of galaxies larger than the Milky Way   inflect downwards towards ``flat'', and RCs of galaxies smaller than the Milky Way   inflect upwards towards ``flat''.  The Milky Way has been  postulated to be at the inflection point  where truly flat RCs lie, though a definitive rotation curve past our position at approximately $8$ kpc has been difficult to obtain until the current era of observations from Gaia DR3 \citep{jiao2023detection}.
We interpret this positioning of the Milky Way as evidence for frame dependent effects in this problem.

 Dark matter theory accounts for the curious URC phenomenology   with the phrasing that galaxies  smaller than the Milky Way   are ``dark matter dominated'', and those larger than the Milky Way  require  only      ``minimal dark matter halos''. Dark matter particles are designed to obey classical gravity, and so this  logical inconsistency  
requires  
fine-tuning  of extra free-parameters,  reducing the predictivity of the models  \citep{MCGAUGH2021220}.
In classical gravity,     mass  accretion rates are  directly proportional to the initial    mass  function  \citep{10.1093/mnras/stt2403}.

In this paper we interpret  the URC phenomenology, instead,  as due to relative frame   effects from our Milky Way, which immediately explains our apparent position in the middle of a spectrum of 1,100 rotation curves. This interpretation of the problem removes extraneous free parameters,   and  does not modify classical gravity theory. 
  Modified Newtonian Dynamics (MOND) \citep{Milgrom}, the leading   alternative     to dark matter, is similar to what we present   in so far as MOND   states that the luminous mass is the only mass, but differs from our model  as it modifies classical gravitational physics   to obtain   rotation curve velocities. We will instead interpret excess rotation velocities, above the estimates from photometry,  as due to an incomplete interpretation of Doppler shifted spectra, not true orbital speeds.

 MOND is instructive in this context because it successfully fits a diverse distribution of galaxy rotation curves (RCs),  by  modifying the standard gravitational acceleration scale $G$ on the length scales of galaxies \citep{McGaugh_2014}. Previous attempts to  extend MOND into the relativistic regime have given     rise to  new theories of gravity  \citep{PhysRevDBekenstein2004,doi:10.1142/S0217751X0703666X,Famaey2012}, but what we propose is  simpler, a transition of the concept of a changing acceleration to the relativistic regime of a changing relative curvature.  It is remarkable  that   
MOND's   free parameter is essentially   constant across a wide range of galaxy morphologies \citep{2016Lelli}, which we interpret as a characterization of the role of the Milky Way in this problem. However, there  are some notable exceptions where MOND fits of  
well-studied galaxies   indicate  distances   which are in conflict with   standard candles, and this problem is removed in our formulation of the flat-rotation curve problem.

The new  rotation curve fitting model (RCFM)   presented here  
  fits  the same  galaxy data as MOND and dark matter models,    at reported distances,  but with fewer free parameters and no modification of  physics. 
  The  RCFM model is a one-parameter fit to the rotation curve data, whose     only inputs are   estimates of the luminous mass of the galaxy emitting the photons and of galaxy receiving those photons.  
 Recent observations of the Milky Way  from the Gaia DR3 \citep{jiao2023detection} indicate a   Keplerian decline in rotation velocities from $8$ kpc to $26$ kpc, consistent with the model paradigm presented here.

 The new rotation curve fitting model (RCFM) is derived in {\bf Section \ref{sec:dos}}. 
  Historically, gravitational redshifts   considered in this problem were  obviated   by Galilean subtraction of redshifts at the large $r$ limit of the data of the emitter galaxy from the  receiver galaxy \citep{MTW}.  We instead employ a    relativistic framing,    in which      field frames are parametrized by the Schwarzschild gravitational redshift terms, and then we Lorentz map between   between the emitter and receiver galaxies   one-to-one in   radius  (See Section ~\ref{sec:RCFM}).    In {\bf  Section \ref{sec:data} } we   describe the   SPARC database \citep{2016Lelli} of 175 well known   galaxy rotation curves  and associated  luminous mass models, and the  Milky Way baryon models used in this work. 
 In   {\bf  Section \ref{sec:analysis} } we 
 compare RCFM fit  results to those from  
  dark matter  models    and the Radial Acceleration Relation  (RAR) \citep{McGaugh_2014,McGaugh2016RAR,2016Lelli,Li_2018} for the SPARC galaxy database (RAR has a simpler functional form and is a possible precursor for   MOND based on the same physical paradigm \citep{10.1093/mnras/stz531}).  
In this section we also  explore   a   correlation between the RCFM  free parameter and a ratio of  photometric observables (luminosity and half light radius of the galaxy).  
 In {\bf   Section \ref{sec:conclu}}  we conclude with  discussion of   upcoming   observations which can further constrain or falsify this model.

\section{ Rotation Curve Fitting Models }
 \label{sec:dos}
 
  \subsection{Two observations of light} 
  The observables in the flat-rotation curve problem are the Doppler shifted spectra and photometry.
  
 Doppler shifted spectra is  interpreted as   
   RC velocities    $ v_{obs}(r)$      by the
 Lorentz boost

 \begin{equation}
 \frac{\omega'(r)}{\omega_o} = \sqrt{\frac{1-\beta(r)}{1+\beta(r)}}, 
\label{eq:modelLumA}
\end{equation} 

for $\omega_o$ the    characteristic lab frequency,  $\omega'(r)$
the observed shifted frequency, and $\beta = v_{obs}(r)/c$, for 
$c$  the  vacuum light speed.  We emphasize this form of the Lorentz boost is applicable to a flat, inertial spacetime.
 
The second observable, photometry,  gives a measure of total light which  is then  interpreted as    mass from population synthesis models (PSMs) which produce  mass-to-light ratios and hence models of the baryon distribution in a given galaxy. 
   PSM rely upon a complex  suite of  assumptions regarding galaxy evolution, metallicities and initial mass functions  \citep{10.1093/mnras/sty3223, BelldYong}, and are under-constrained   due to the dark matter problem \citep{Conroy,Dutton_2005}, producing an  error budget of  $\approx 20\%$ \citep{2016Lelli}.
 
 Mass models give velocities
 by parametrizing   the  integration of the Newtonian gravitational potential  by
   
\begin{equation}
      \Phi(r)  = - \int_{r_{ref}}^{r}   \vec{g(r)}\cdot\vec{dr} .
      \label{eq:Newt}
\end{equation}
  
for  $r$ the field point,  $r_{ref}$ the reference point, and  $g$   the acceleration due to the   gravitational field. 
The classical boundary condition is that the potential goes to zero at $r'=\infty$, hence the integration effectively starts from large $r_{ref}=R$ limit of the data, and integrates in to the small $r$.  
This potential then solves the Poisson equation
  
\begin{equation}
    \nabla^2 \Phi(r)_{lum}  = 4\pi G \rho(r'),     
    \label{whatsgood}
\end{equation}
 
where $\rho(r')$ is the mass density distribution and $G$ is Newton's constant.
The gradient of this potential then   gives the circular orbital velocities $v(r)_{lum}$ (Eq.~\ref{eq:zonte1}) by the central force relation

\begin{equation}
 \frac{\partial \Phi(r)_{lum}}{\partial r}    =\frac{v(r)_{lum}^2}{r}. 
    \label{zoochance1}
\end{equation}

 The  
  baryonic components of a galaxy include the    stellar disk,  stellar bulge,   and a gas halo, which are   summed in quadrature to reflect a sum of gradients
  
   \begin{equation}
v(r)_{lum}^2 = \gamma_b v(r)_{bulge}^2 +  \gamma_d v(r)_{disk}^2 + v(r)_{gas}^2.  
\label{eq:zonte3}
\end{equation} 
  
Mass-to-light ratios $\gamma_b$ and $\gamma_d$   are   free parameters  in  most rotation curve fits due to the under-constrained nature of luminous mass modeling.
  Gas fractions $v(r)_{gas}$  are calculated from a different observational technique  \citep{1983MNRAS.203..735C}, and do not require  mass-to-light ratio modeling.

 \subsection{Dark matter rotation curve fitting formula}
 
 The    dark matter rotation curve (RC) formula    is  formulated in the same sense as Eq.~\ref{eq:zonte3}, as a sum of gradients in the potential

 \begin{equation}
v(r)^2_{rot}  =  v(r)^2_{lum}  +  v(r)^2_{dm},   
\label{eq:zonte1}
\end{equation} 

  for terms in
  $v(r)_{lum}$   the Keplerian velocity predictions (Eq.~\ref{eq:zonte3}),  
  $v(r)_{dm}$   the velocities attributed to  dark matter, 
  and  
  $v(r)_{rot} $   the model  predictions which are fitted  to the RC velocities $v(r)_{obs} $ reported from observations of  Doppler shifted  spectra (Eq.~\ref{eq:modelLumA}). 
  All  velocities are assumed to be those of test particles in circular orbits, in the plane of the stellar disk,  about the rotation axis of the galaxy at  $r=0$.

\subsection{Radial Acceleration Relation }

 The Radial Acceleration Relation (RAR)\citep{McGaugh2016RAR}   has been considered as the  
   phenomenological basis for MOND \citep{10.1093/mnras/stz531,10.1093/mnras/stad597}, as it has a simpler functional form and less scatter than MOND but the same physical paradigm. For this reason, we will  compare RCFM fits to 
     RAR   and dark matter fits to the SPARC sample. 
   The RAR
       approaches   rotation curves from  the    MONDian paradigm of a changing law of inertia, but has also been shown to be consistent with dark matter fits on a sample of simulated galaxies \citep{Keller_2017}.

   


\subsection{ New rotation curve fitting model  (RCFM)   }
\label{sec:RCFM}

The physical paradigm  for   the  new rotation curve fitting model (RCFM)  is:  the luminous mass is  the only mass,  the velocity $v_{lum}$ due to the baryons 
is the only velocity, and   
   excesses in Doppler shifted spectra represented by the ``flat'' rotation velocities $v_{obs}$ are due to relative frame effects from  the Milky Way galaxy.
    For clarity in what follows, we   now    drop       explicit   functional dependence on $r$ and set constants  like Newton's   $G$ and the speed of light $c$  to 1.
    All terms can be assumed to be evaluated  one-to-one in radius between the galaxy being observed and the Milky Way, with the exception of    the model's free parameter $\alpha$ which is single valued for each galaxy fitted.  
   
   We construct a new heuristic  rotation curve fitting formula by replacing  the dark matter 
   contribution $v^2_{dm}$  to  the rotation curve formula in  Eq.~\ref{eq:zonte1}  with the convolution of two Lorentz-type maps  $S_{1}$ and $S_{2}$ which represent the frame-dependent  effects due to the Milky Way;
  
 \begin{equation}
v_{rc}^2 =  v_{lum}^2+\alpha \kappa^2 S_{1} S_{2}.   
\label{eq:zonteLCM}
\end{equation}   

 This prescription is an     implicit   assumption that  contributions to  Doppler shifted spectra from  translation  ($v_{lum}$  in Eq.~\ref{eq:zonte3}) and      curvature  ($S_{1} S_{2}$)   are   roughly separable \citep{Jack} as represented by Eq.~\ref{eq:zonteLCM}.

 Terms in    $\kappa$
are a measure of relative curvature 

 \begin{equation}
\kappa=\frac{\Phi_{gal}}{\Phi_{mw}}. 
\label{eq:kappa2}  
\end{equation}

The   RCFM galaxy     maps $S_1$ and $S_2$   (Eq.~\ref{eq:zonteLCM})    are constructed  from  the tetrad formalism     of general relativity \citep{universe5100206},  where an  inertial frame field metric  (Minkowski $\eta^{a b}$)
  can be attached to any point on a curved manifold.
The frame fields $e^\mu_a  $ are related to the curved spacetime metric $g^{\mu \nu}$ 
in the following way

\begin{equation}
    g^{\mu \nu} = e^\mu_a e^\nu_b \eta^{ab}.
\end{equation}

 In  the frame field  formalism     ``when going from
one local inertial frame at a given point to another at the same point, the fields transform
with respect to a Lorentz transformation'' \citep{jetzer}.
Our goal  then, is to  use  Lorentz-type transformations to  map target  galaxies onto the Milky Way, one-to-one in radius. To do this, we use   the  weak-field Schwarzschild time metric coefficients $g_{00} =-(1-2\Phi )$ to characterize the frame fields, and hence the effects on photon frequency. 
The   weak-field timelike  Schwarzschild field frames are defined by
 
 \begin{equation}
     e^0_0  =\sqrt{1-2\Phi}.
     \label{eq:coframe}
 \end{equation}
 
 Terms in  $\Phi$ are
 the Newtonian gravitational potential \citep{Wald}. 
Classically, the  $\Phi$   of galaxies are integrated  from the   large  $R$ limit of the data with  a boundary value of $\Phi = 0$, 
 into a negative maxima    at the small $r$ limit (Eq.~\ref{eq:Newt}).
This   is an implicit assumption of  a  globally  flat embedding space, which   ensures   that  the potential goes to zero at a distance of infinity from the mass distribution. 

However, recent  observations \citep{Pomarede:2020pme}  have   shown that on the  relevant  length scales   of galaxies and groups of galaxies we have not recovered flat spacetimes. 
In fact, the external environments of galaxies  and flowlines of groups of galaxies    are exceedingly  complex and diffusely populated with matter, demonstrating structure with  sources and sinks.  In addition, the value of the vacuum energy remains an outstanding physics problem \citep{YaBZel'dovich_1968}.

In the absence of definite  knowledge of the   values of $\Phi$ at the large $r$ limit of individual galaxies, 
and because all RCFM terms are ratios of terms in $g_{00}$ for the emitter and receiver galaxies, we instead  want to compare galaxies from a position where we know they have the same curvature, namely at their centers. To inform the integration of the gravitational potentials, we note   Wolfgang Rindler's statement that    ``the center of each galaxy provides a basic local standard of nonacceleration ... so then can be treated like a local inertial frame relative to its own center''\citep{rindler2013essential}.
 From Rindler's insight,  we  propose that to compare galaxies as inertial frames by Lorentz-type maps we must integrate  from their centers out to the limit of the data. 
In practice, this means that galaxy potentials are summed from   $\Phi=0$ at  $r\approx 0$ out to  an essentially constant  value
at the large $R$ limit of the data. This  produces positive definite gravitational potentials, 
    which are    then   subtracted from $1$ in  the Schwarzschild clock term $g_{00}=-(1-2\Phi)$. 
Gravitational 
potentials calculated in this way still  obey Poisson's equation Eq.~\ref{whatsgood} and the central force law Eq.~\ref{zoochance1}, but allow for a non-flat embedding space with no loss of generality.  \\

\begin{figure}
	\includegraphics[width=\columnwidth]{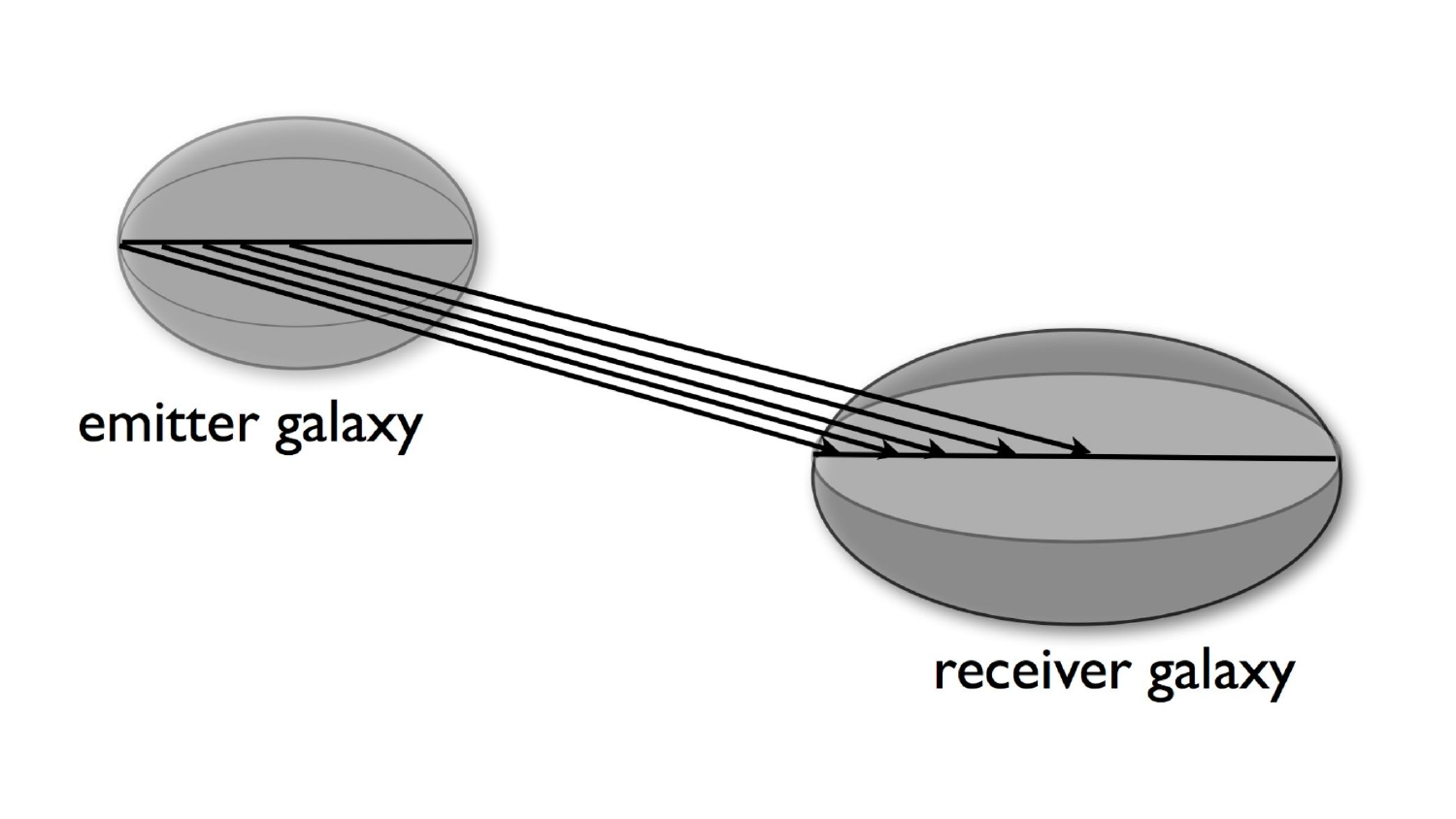}
    \caption{Cartoon of the $S_1$ transformation 
 of curved 2-frames. Image credit: N. Oblath  }
   \label{NoahCartoon1}
\end{figure}

The RCFM mapping is accomplished in two steps. First, the  curved 2-frame mapping    $S_1$  maps the entire emitter galaxy onto the Milky Way one-to-one in radius, 

 \begin{equation}
       S_1 = \sinh \zeta, 
       \label{eq:hyperbolica}
   \end{equation}
   
 for   a rapidity defined by   
   
    \begin{equation}
     e^{\zeta}=  \sqrt{\frac{g_{00}|_{gal}}{g_{00}|_{mw}}}, 
      \label{eq:gravRS}
    \end{equation}
    
 where $g_{00}|_{gal}$ represents  the emitter galaxy   and $g_{00}|_{mw}$   the Milky Way. A cartoon of the $S_1$ transformation is represented in Fig.~\ref{NoahCartoon1}.

 The  second transformation takes this curved 2-frame (Eq.~ \ref{eq:gravRS})  into the flat field frames where observations are made and where   the metric is   Minkowski $\eta^{\alpha \beta}$, as    

\begin{equation}
S_{2} =  \cosh \tau  
\label{eq:hyperbolico}
\end{equation}

  for a rapidity angle $\tau$ defined by

\begin{equation}
    e^{\tau}=   e^{(\zeta+\eta)},
\end{equation}
 
for    $e^{\tau}$ a    convolution   of the curved  frame $e^\zeta$ (Eq.~ \ref{eq:gravRS})  with  the flat  frame  $e^\eta$

\begin{equation}
    e^{\eta}=  \sqrt{\frac{1+\beta'}{1-\beta'}}.
    \label{eq:flat}
\end{equation}

\begin{figure}
    \includegraphics[width=\columnwidth]{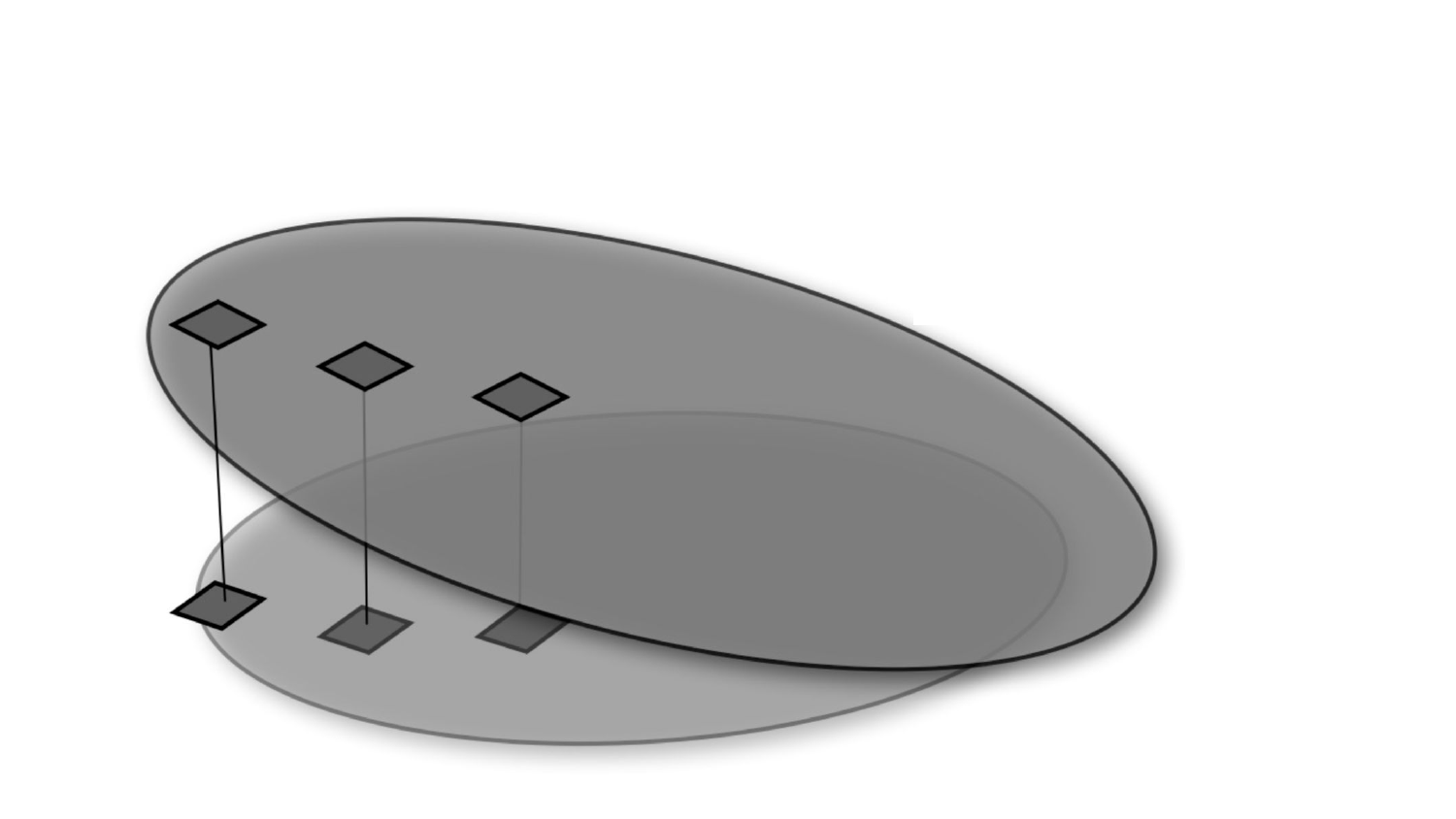}
    \caption{Cartoon of $S_2$    mapping from curved 2-frame to flat 2-frame. Image credit: N. Oblath   }
  \label{NoahCartoon2}
\end{figure}

Terms in    $\beta' = v_{lum}/c$  are the  Keplerian velocities  from luminous mass as in Eq.~\ref{eq:zonte3}.   A cartoon of the $S_2$ transformation is represented in Fig.~\ref{NoahCartoon2} .

That the   Keplerian velocities  are  our best  estimate of flatness is evidenced by the fact that  dark matter is not required to  reproduce the RC of our Solar System.  

Previously, Rindler extended the     transforms of special relativity  
to     accelerated frames on  a flat background  \citep{2011AmJPh..79..644B}.  
   Viewed  as Rindler's accelerated coordinates, $S_1$  and $S_2$ respectively can be seen to be   timelike   and spacelike transformations \citep{MTW,Wald, rindler2013essential}.

\subsection{  Geometric simplifications  }\label{GeomSphere}
  
   Population synthesis models commonly assume spherical symmetry for the 
  stellar bulge and gas halo, but   axial symmetry for    the stellar disk \citep{1954AJ.....59..273S,Freeman}.
 However, it is a common calculational  tool to use spherical symmetry for the entire integrated potential of the luminous mass distribution  \citep{PhysRevDBekenstein2004, McGaugh_2008, 2022A&A...664A..40M}  because numerical integration of the thin disk is  computationally intensive,  requiring  assumptions of under-constrained boundary conditions and  relevant physical scales,  which therefore add extra free parameters to the problem \citep{2011A&A...531A..36H}.  
 We   use the spherically symmetric Schwarzschild metric here as proof of concept, as it is  the simplest   representation  which captures the relevant physics of the problem without excess computations.

For a given mass, the gravitational potentials  calculated in  a spherically symmetric  geometry    converge to those   for an exponential disk geometry  at lengths greater than one-third of the exponential scale-length   $R_e$, ie.  $r> R_e/3$  \citep{Chatterjee}.  
  Since this is the region where   dark matter effects becomes important \citep{1985ApJAlbada},  this calculational technique captures the relevant physics.
  However, in the region where the RCFM integrations begin galaxy comparisons   ($r\approx 0$ to $R_e/3$)   
  the spherical assumption   
    overestimates the potential by a factor of $\approx  2$, which can   be seen in RCFM disk mass-to-light ratio results.   Implementation of a thin   disk geometry would resolve this problem \citep{1959HDP....53..275D},\citep{Freeman}, though at significant computational expense.

\section{Data }
\label{sec:data}
 
 \subsection{SPARC galaxies}
 We fit the Spitzer Photometry and Accurate Rotation Curves (SPARC) dataset  of  175 nearby galaxies with extended RC data from atomic hydrogen (HI)  and H-$\alpha$ \citep{2016Lelli}. 
 HI  provides the most reliable
 RCs because it is dynamically cold, traces circular orbits, and can be observed several effective radii past the stellar disk. 
 This sample of rotationally supported galaxies   spans the widest range of masses and morphologies presently available. 
 
These galaxies are  accompanied by Keplerian velocities which represent the baryonic mass model  from PSM, based on 
   Spitzer Photometry in the 
   near infrared  at 3.6$\mu m$.
   Near infrared  is  currently believed to be the best tracer of stellar mass   in population synthesis models (PSM) \citep{10.1093/mnras/sty3223}, as at this wavelength,     mass-to-light ($\gamma_i$) ratios   are believed to be almost constant and  independent of star formation history \citep{BelldYong,10.1093/mnras/sty3223}. 
  The SPARC database  reports     mass-to-light ratios of   $\gamma_i=1$ in units of $M_{\odot} / L_{\odot}$   at 3.6$\mu m$. 
 Gas fractions $v_{gas}$ are calculated from surface density profiles of HI   with the formalism given in  \citep{1983MNRAS.203..735C} and scaled by
     a factor 1.33 to account for cosmological helium abundances.  Contributions from molecular gas are ignored   because CO data are not available for most SPARC galaxies. 
     Error on these velocities is estimated at $20\%$ \citep{2016Lelli}. The      SPARC  database can be found  at \url{http://astroweb.cwru.edu/SPARC/}.

\subsection{Milky Way Luminous Mass Models}
\label{MWselect}

  The RCFM,      as currently formulated,   requires a static choice of a   luminous mass distribution for the  Milky Way. 
   Determining the luminous mass profile of  the Milky Way (MW) is an  under-constrained problem, due to  our   observing position from       within  the galactic disk  \citep{1991ARA&A..29..409F}. 
  In this paper, we        compare two different   Milky Way  baryon models to the SPARC sample; one from  McGaugh \citep{McGaugh_2008} and one from  Sofue \citep{Sofue}. 
  
  The  two MW models differ markedly in the inner   $7$ kpc; as the McGaugh MW has a triaxial bar-bulge  and the Sofue MW has a de Vaucouleurs bulge. 
   The Sofue MW model 
  is used from  $0$ to $20$ kpc,   and extended with  the Xue MW model from $20$ to $60$ kpc \citep{Xue}.  The Sofue and Xue MW models come   from the same data and a   dark matter  model  is used for the extended rotation curve.   The McGaugh Milky Way model covers from $0$ to $150$ kpc, comes from data, and the  MOND model is  used for the extended rotation curve.

  The  recent ESA \emph{Gaia}
   data release (DR3),  \cite{jiao2023detection} demonstrates   a Keplerian decline of the rotation curve   of the Milky Way in the range of  $9.5$ to $26.5$ kpc, consistent with this RCFM heuristic paradigm.  
    The ESA \emph{Gaia} mission, taking measurements from  the second Lagrange point,  has revolutionized the science of the   MW with
     unprecedented detail,  statistical accuracy,  and a drastic reduction in  systematic uncertainties.   Since this is the most recent and best data for the MW, we  replace  the model Milky Way model velocities in the range from $9.5$ to $26.5$ kpc with those from    \cite{jiao2023detection}. We  then shift the two remaining sides of the given MW  model     by a constant amount  to match   the endpoints of the Jiao rotation curve. See Table \ref{MW_dats} for summary details.

  \begin{table*}	
\centering				
\caption{ {\bf Milky Way Models} \label{MW_dats} 
\emph{For both   models, the kinematic representation of the luminous mass is    replaced in the region from $9.5$ to $26.5$ kpc with the recent Gaia DR3 velocities \citep{jiao2023detection}.}}				
\begin{tabular}{|c|c|c|c| }				
\hline				
Author&	 	scale length&Model  for the &	Range\\
    &	   bulge/bar  &	extended RC   &	\\
\hline				
Sofue-Xue  	&0.5 kpc     &	NFW  \citep{1996ApJ...462..563N}  	&[0,60] kpc \\
\cite{Sofue,Xue}&  bulge & dark matter &	 	\\
\hline				
McGaugh &	 	2.0 kpc&	MOND   	&[0,150] kpc \\
\cite{McGaugh_2008} &bar&   \citep{Milgrom}    &	\\
\hline				
\end{tabular}	
 \end{table*}

\section{  Analysis and Results } \label{sec:analysis}
 
\subsection{Fitting procedure}
 
To fit galaxies in the SPARC sample with the RCFM model,  the fitting procedure is as follows.  
First,  a       Milky Way baryon model is  selected (Sec.~\ref{MWselect})
and the model data in $v_{lum}$ is  read in  for a series of measurements in radii. The galactic gravitational potential is calculated by numerically integratingas   (see  Sec.~\ref{sec:RCFM}), 

\begin{equation}
\Phi(r) = \int_{inner}^{outer} dr \frac{ 
v(r)^2_{lum}
}{r}.
\label{nowthen}
\end{equation}

Once the MW potential is calculated, it remains static for the rest of the fitting procedure.  

The data for the   galaxies being observed include several pieces of information: the RC velocities $v_{obs}$ from Doppler shifted spectra, the uncertainty on that measurement $v_{err}$, and the components of the luminous mass interpreted as orbital velocities,  $v_{bulge}$, $v_{disk}$, and $v_{gas}$. 
To calculate the baryonic potential for the galaxy in question, $v_{lum}$ is first computed   as per Eq.~\ref{eq:zonte3}, for $\gamma_b=1$ and $\gamma_d=1$. The $\Phi_{gal}$ associated with that galaxy   is then computed as in Eq.~\ref{nowthen}.

After the $\Phi$ for the galaxy being studied and the Milky Way have been calculated,   the components must be compared at  matching values of $r$. To match radii, $\Phi_{MW}$ is interpolated to produce values at the   radii reported in the measured 
RC data $v_{obs}$ of the galaxy being observed. Any point with a radius larger than the largest radius in the Milky Way model is discarded.

 The RCFM prediction is    assembled as in Eq.~\ref{eq:zonteLCM}, to give a predicted $v_{rc}$ which is compared to the RC data $v_{obs}$. 
The equations outlined in Section \ref{sec:dos} contain   free parameters that must be determined for each galaxy fit: $\alpha$, $\gamma_b$, and $\gamma_d$. The model's free parameter $\alpha$ starts from  an initial value of $\alpha = 0.01$, and  mass-to-light parameters from an initial value of $1.00$, and then all parameters are allowed to vary freely and to be determined by  minimization of the $\chi^2$.  The {\tt scipy.optimize.curve\_fit} utility in Python is used to perform this minimization.

 As noted in Section \ref{sec:dos}, reported gas fractions  are fixed at their reported values (HI scaled for Helium abundance by a factor of 1.33 in the SPARC database),  though addition of molecular gas could increase mass fractions in the inner kiloparsec of a galaxy   \citep{2004ApJ...609..652M}.

\subsection{Evaluating goodness-of-fits}

There are two metrics by which different rotation curve fitting models are compared, the resulting  reduced  $\chi^2_r$ values and 
 the   mass-to-light ratios. 
 Since   error    estimates on RC velocities  have not been standardized across the field~\citep{Blok,Gent},     $\chi^2_r$ values can   only be compared when   fitted to the same RC data. 
 In Table~\ref{table:M2Light} we compare  $\chi^{2}_r$ values from three models:   dark matter halo, Radial-Acceleration-Relation (RAR) \citep{McGaugh2016RAR,Lelli_2017,Li_2018} and 
 RCFM, fitted to the     SPARC database of galaxy RCs \citep{2016Lelli} .   
 RCFM $\chi^2_r$ values are remarkably low, providing confidence in the faithfulness of the model to galaxy RC data. 
 
   The  average mass-to-light ratios from   RCFM fits yield bulge  mass-to-light ratios which are within the error estimates on population synthesis models (PSM). The RCFM disk  mass-to-light ratios are approximately a factor of two larger than PSM, consistent with the  artifact introduced from the 
  spherical integration technique, see
  Sec.~\ref{GeomSphere} for a full explanation. This artifact can be removed by including an exponential disk  integration at the cost of additional computation.

   \subsection{Comparing Milky Way models}

   RCFM fits, as currently formulated,  require a choice of a static  Milky Way baryon distribution. In  Table \ref{tab:lobes} and \ref{TSet} we compare the RCFM fit results for the two 
      different    Milky Way model assumptions used in this paper. As can be seen in the table, for the 36 most reliable galaxies in the SPARC sample the two Milky Way models are equivalent (see Sec.~\ref{FreeCorrel}). However, for the whole SPARC sample of 175 galaxies the RCFM fits using the Sofue-Xue-Jiao Milky Way performs better with  an average $\chi^2_{r}=2.39$, versus those from McGaugh-Jiao Milky Way $\chi^2_{r}=7.09$.
   Based on this,   values  in 
  Table~\ref{table:M2Light} and    figures  are from RCFM fits assuming the  Sofue-Xue-Jiao Milky Way.

\begin{figure}
\begin{subfigure}{.5\textwidth}
  \centering
  \includegraphics[width=.8\linewidth]{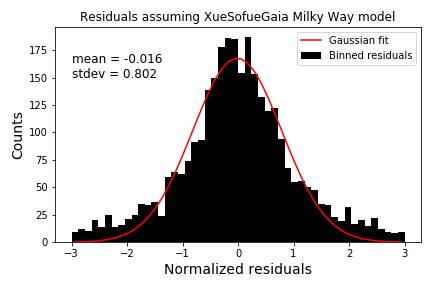}
\end{subfigure}\\
\begin{subfigure}{.5\textwidth}
  \centering
  \includegraphics[width=.8\linewidth]{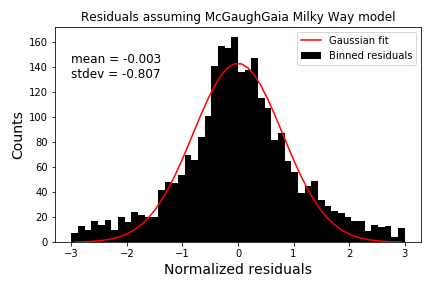}
\end{subfigure}
\caption{Normalized residuals from fits assuming either Sofue-Xue-Jiao or McGaugh-Jiao Milky Way models, fitted by a Gaussian  function. The means and standard deviations are shown.}
\label{fig:residualgraphs}
\end{figure}

    The residuals of fits to the SPARC sample are also used to compare the two different MW models (see Sec.~\ref{MWselect}). 
    Histograms of residuals normalized by the error in velocity observations are shown in Fig.~\ref{fig:residualgraphs}. 
     In all cases, residuals of model fits to observed velocity data followed a narrow distribution centered at zero with a range of $\pm 3 $ standard deviations in error, albeit with heavy tail features. The behavior of the residuals did not vary greatly between MW models, suggesting that the fitting parameters in our model are robust with respect to differing MW model assumptions at this present level of analysis.  With a finer grained analysis and a larger sample, the degeneracy between MW models may be resolved. 
    
    A Gaussian fit on residuals from fits using the McGaugh-Jiao MW model gave a mean of $-0.003$, standard deviation of $0.807$ in units of standard deviations in velocity error; whereas a Gaussian fit to the residuals from fits using the Sofue-Xue-Jiao MW model gave a mean of $-0.016$, standard deviation of $0.802$ in those same units. The residuals are shown in Fig.~\ref{fig:residualgraphs}. The small values associated with these quantities in both cases provide confidence that our fits match data closely.

\begin{figure}
\begin{subfigure}{.5\textwidth}
  \centering
  \includegraphics[width=.8\linewidth]{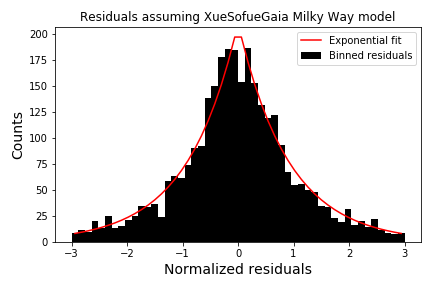}
\end{subfigure}\\
\begin{subfigure}{.5\textwidth}
  \centering
  \includegraphics[width=.8\linewidth]{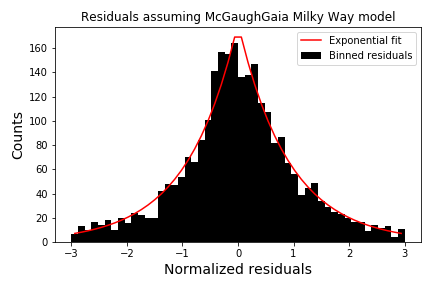}
\end{subfigure}
\caption{Normalized residuals from fits assuming either Sofue-Xue-Jiao or McGaugh-Jiao Milky Way models, fitted by an exponential function. }
\label{fig:EXPresidualgraphs} 
\end{figure}

However, the Gaussian fits did not quite capture the full peak and heavy tails in the residuals. This suggests that there may be non-Gaussian error in the observations of galaxy velocities. To address this, the residuals were also fit to an exponential function shown in Fig.~\ref{fig:EXPresidualgraphs}. The exponential function captured both the peak and the heavy-tailed behavior of the distributions more faithfully.

 \subsection{  Individual Galaxy Results}
 \label{results:MtoL}

   In this section we will  
   compare fit results from the RCFM, RAR
   and dark matter models for 
      well-studied galaxies. All reported $\gamma_i$ in what follows are in units of  solar mass per solar luminosity $M_\odot/L_\odot$.

\subsubsection{NGC 2841}
 
\begin{figure} 
\begin{subfigure}{.5\textwidth}
  \centering
  \includegraphics[width=.8\linewidth]{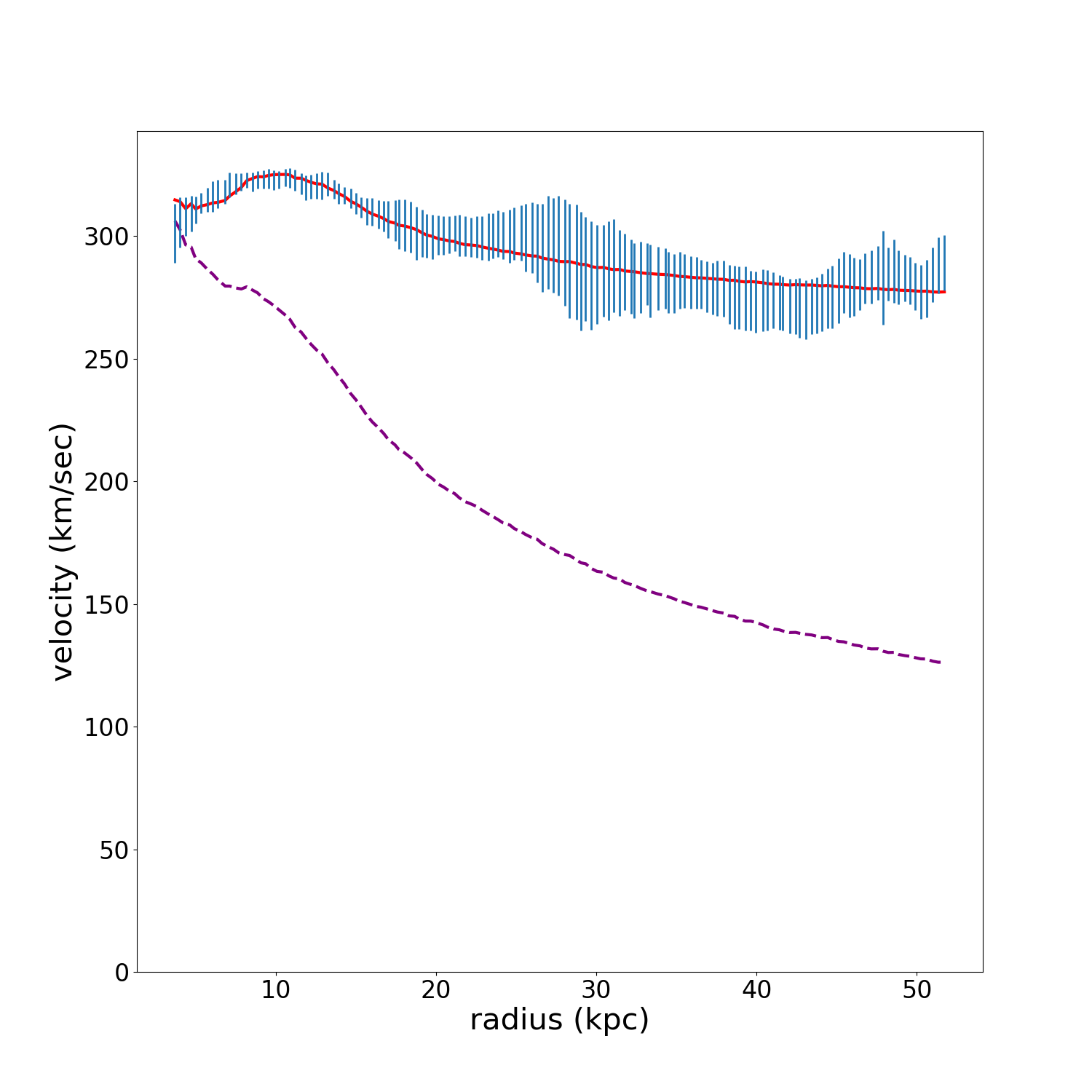}
  \caption{NGC 2841 RC data and baryonic mass model from  \citep{Blok1}}
\end{subfigure}\\
\begin{subfigure}{.5\textwidth}
  \centering
  \includegraphics[width=.8\linewidth]{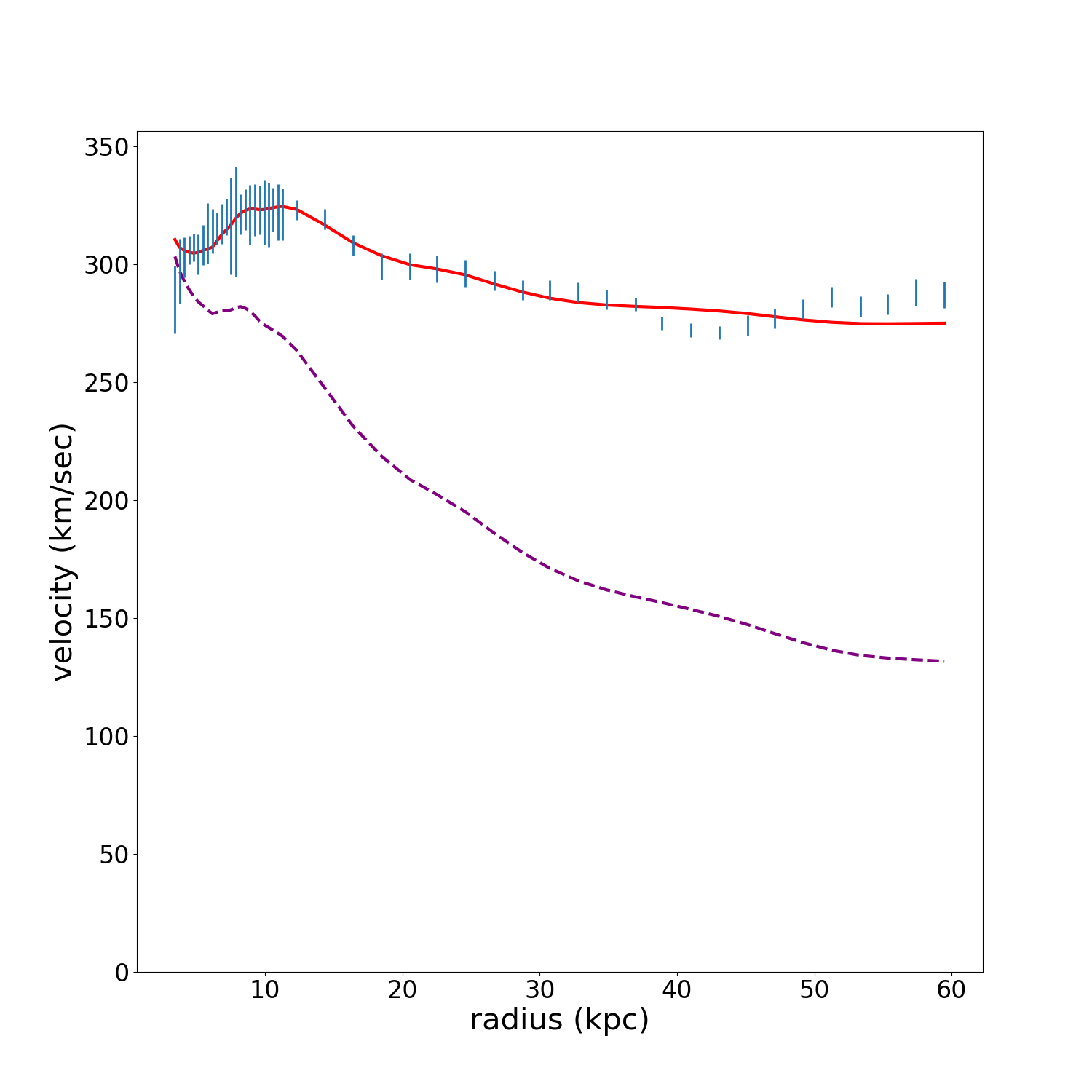}
  \caption{NGC 2841 RC data and baryonic mass model from \citep{2016Lelli}.}
\end{subfigure}
\caption{NGC 2841 RCFM fits.  
 Lines are as in Fig.~\ref{fig:NGC2403RCFM}.}
\label{fig:CompareNGC2841} 
\end{figure}

NGC 2841 is a star dominated, flocculent spiral galaxy, which   historically has been regarded as a problematic case for MOND  (\citep{Gent}).   RAR   finds a good  fit for this galaxy    after adjusting
the Cepheid   distance of $14.1$ Mpc  by $1\sigma$,  to $15.5$ Mpc.   Cepheid variable stars are the most accurate  distance indicators currently available.
The RCFM fit is shown in   Fig.~\ref{fig:CompareNGC2841} at the Cepheid based distance, and fit results are compared to dark matter and RAR in Table ~\ref{tab:N2841}.

 \begin{table} 
        \centering
     \caption{  NGC 2841. Results from fits to the SPARC    data and luminous mass model. \protect\cite{2016Lelli}} 
        \begin{tabular}{|c|c|c|c|c|c|}
       \hline 
        \hline
        & Model             & $\gamma_d$    & $\gamma_b$    & $\chi^2_R$ & distance(Mpc) \\
        \hline
 \hline
        & Dark Matter Iso   &0.60           &     0.66      & 1.58          &  14.1\\
\hline
         &  RAR	       &   0.81        &	 0.93      &1.52          &15.5 \\
\hline
         &  RCFM           &  0.91           &1.10          &1.36           & 14.1  \\
\hline
        \end{tabular}
         \label{tab:N2841}
    \end{table}

\subsubsection{IC 2574}

\begin{figure}
\begin{subfigure}{.5\textwidth}
  \centering
  \includegraphics[width=.8\linewidth]{ 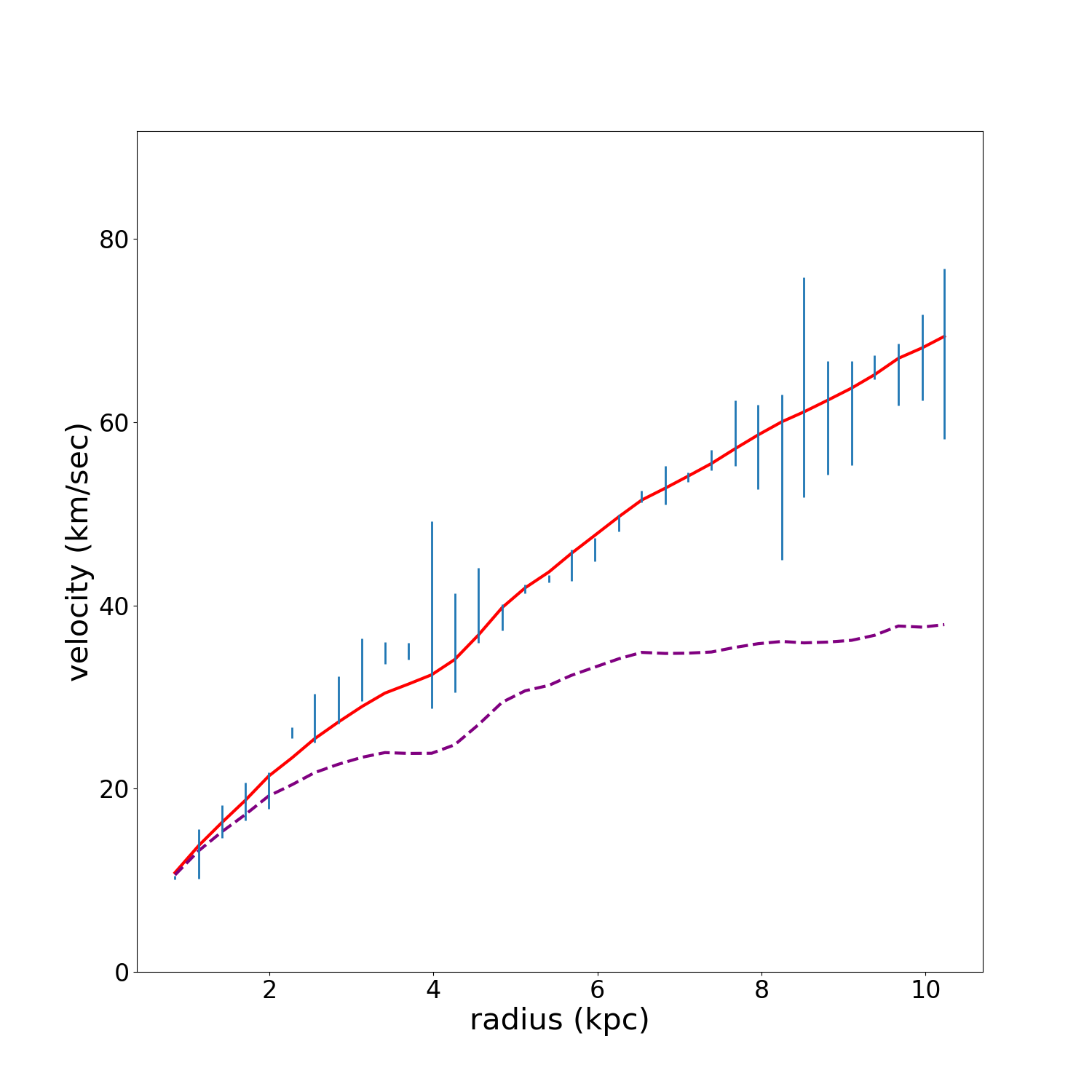}
  \caption{\text{IC 2574}  }
\end{subfigure}\\
\begin{subfigure}{.5\textwidth}
  \centering
  \includegraphics[width=.8\linewidth]{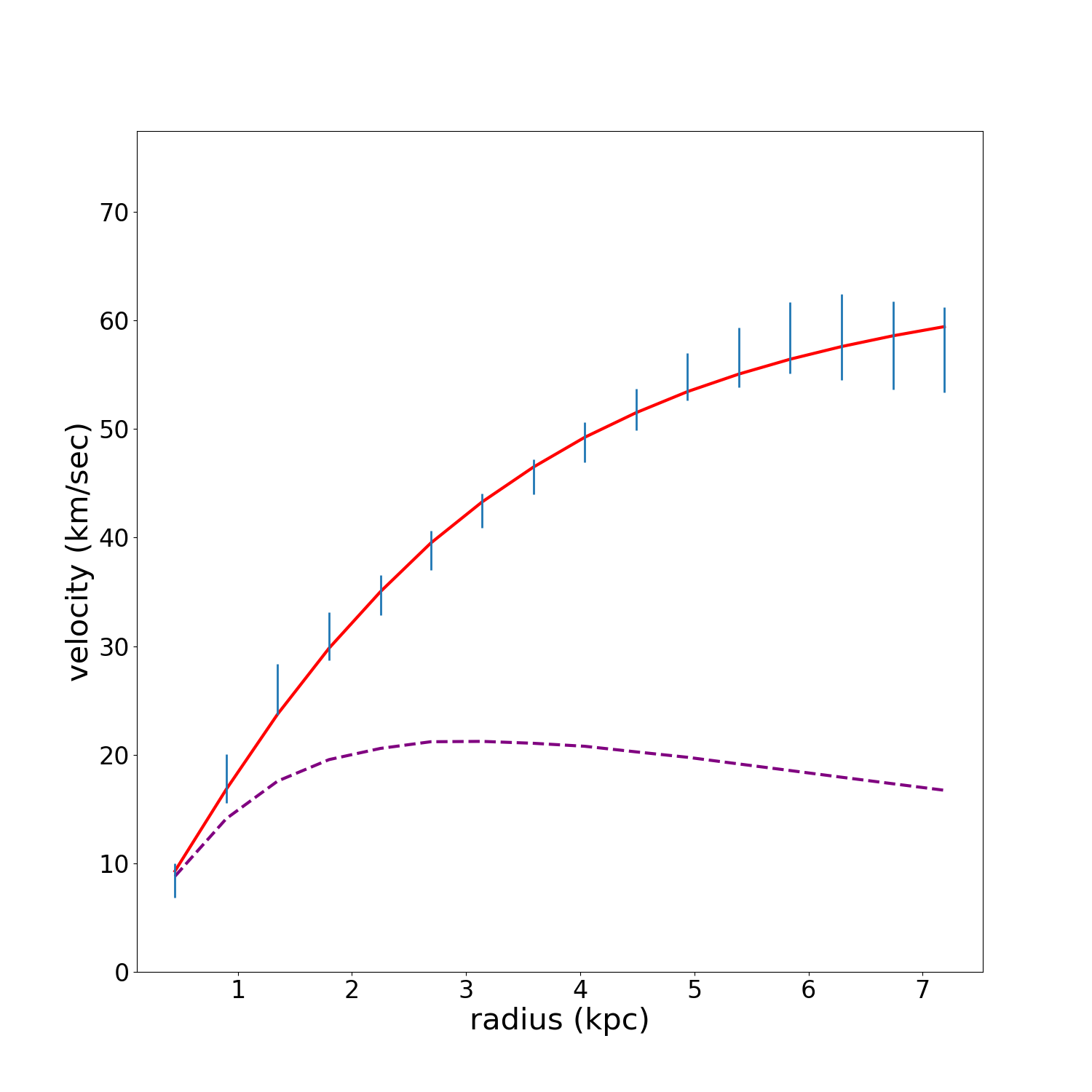}
  \caption{ \text{D 631-7}   }
\end{subfigure}
\caption{   RCFM fits  to gas dominated dwarf galaxies.    RC and baryon models from \citep{2016Lelli}.    Lines are as in Fig.~\ref{fig:NGC2403RCFM}.  }
\label{fig:CompareIC2574}
\end{figure}

IC 2574 is  a dwarf spiral galaxy,    dominated by gas with no central stellar bulge.
 RAR finds a good fit to this galaxy    but adjusts
the  distance and inclination by $1$ to $1.5 \sigma$.
This galaxy is   problematic for dark matter model fits, as the   model   overestimates  the inner RC out to 10 kpc  \citep{2017MNRAS.471.1841N}.  
RCFM fits this galaxy successfully, at the reported  tip of the red giant branch distance  of  $3.91$ Mpc, Fig.\ref{fig:CompareIC2574}, model comparisons are in Table~\ref{tab:IC2574}.
 
 \begin{table} 
        \centering
        \caption{ IC 2574. Results from fits to the SPARC    data and luminous mass model. \protect\cite{2016Lelli} } 
        \begin{tabular}{|c|c|c|c|c|}
       \hline 
        \hline
        & Model            & $\gamma_d$ & $\gamma_b$   & $\chi^2_r$   \\
        \hline
 \hline
        &Dark Matter Iso   &  0.77    &    --     & 2.51     \\
\hline  
         &RAR	       &  0.07   &	 --     &  1.44    \\
\hline
         &RCFM              &1.10     &    --        &   2.27    \\
\hline
        \end{tabular}
         \label{tab:IC2574}
    \end{table} 

\subsubsection{NGC 3198}
 
\begin{figure} 
\begin{subfigure}{.5\textwidth}
  \centering
  \includegraphics[width=.8\linewidth]{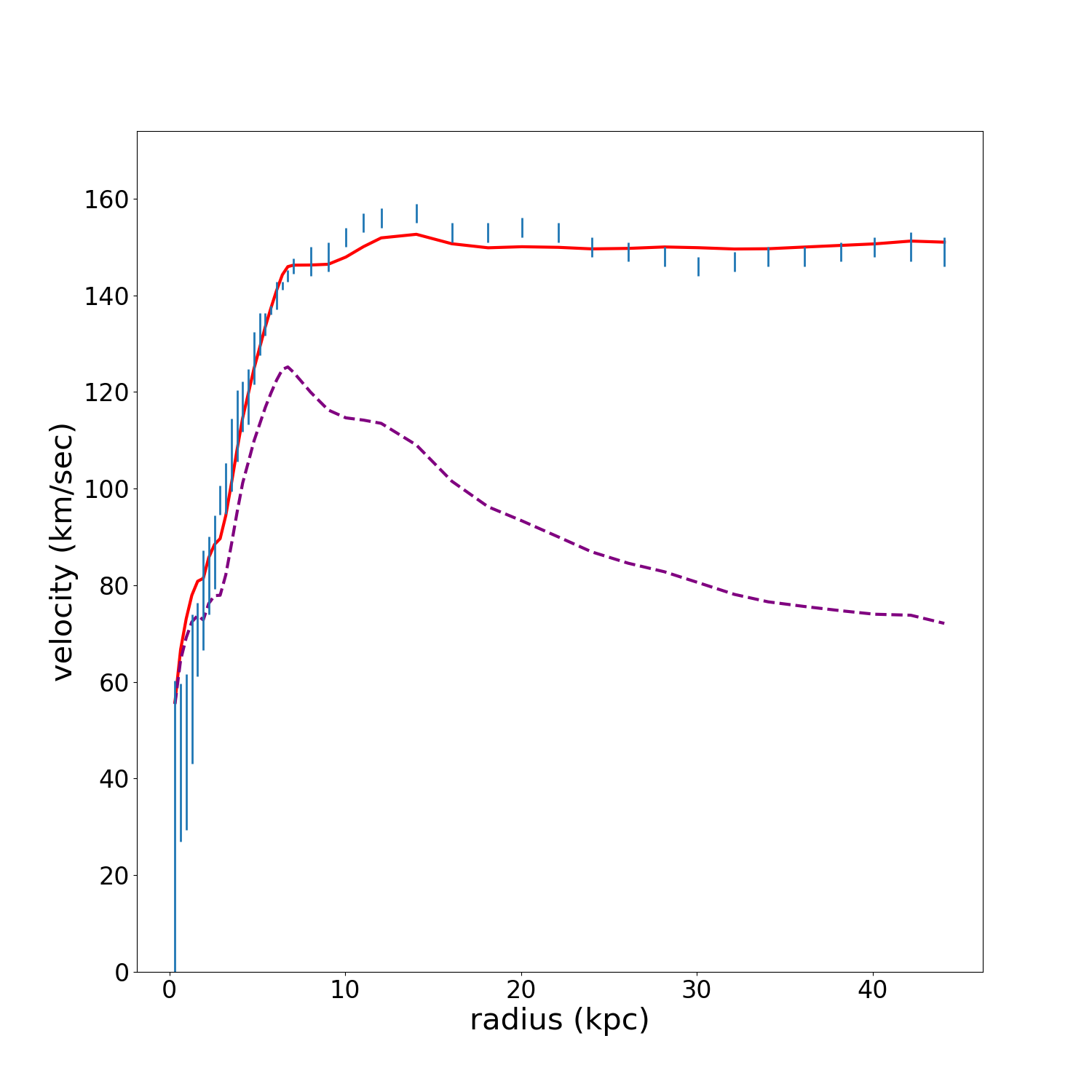}
  \caption{NGC 3198, RC and baryon model from \citep{2016Lelli}. }
\end{subfigure}\\
\begin{subfigure}{.5\textwidth}
  \centering
  \includegraphics[width=.8\linewidth]{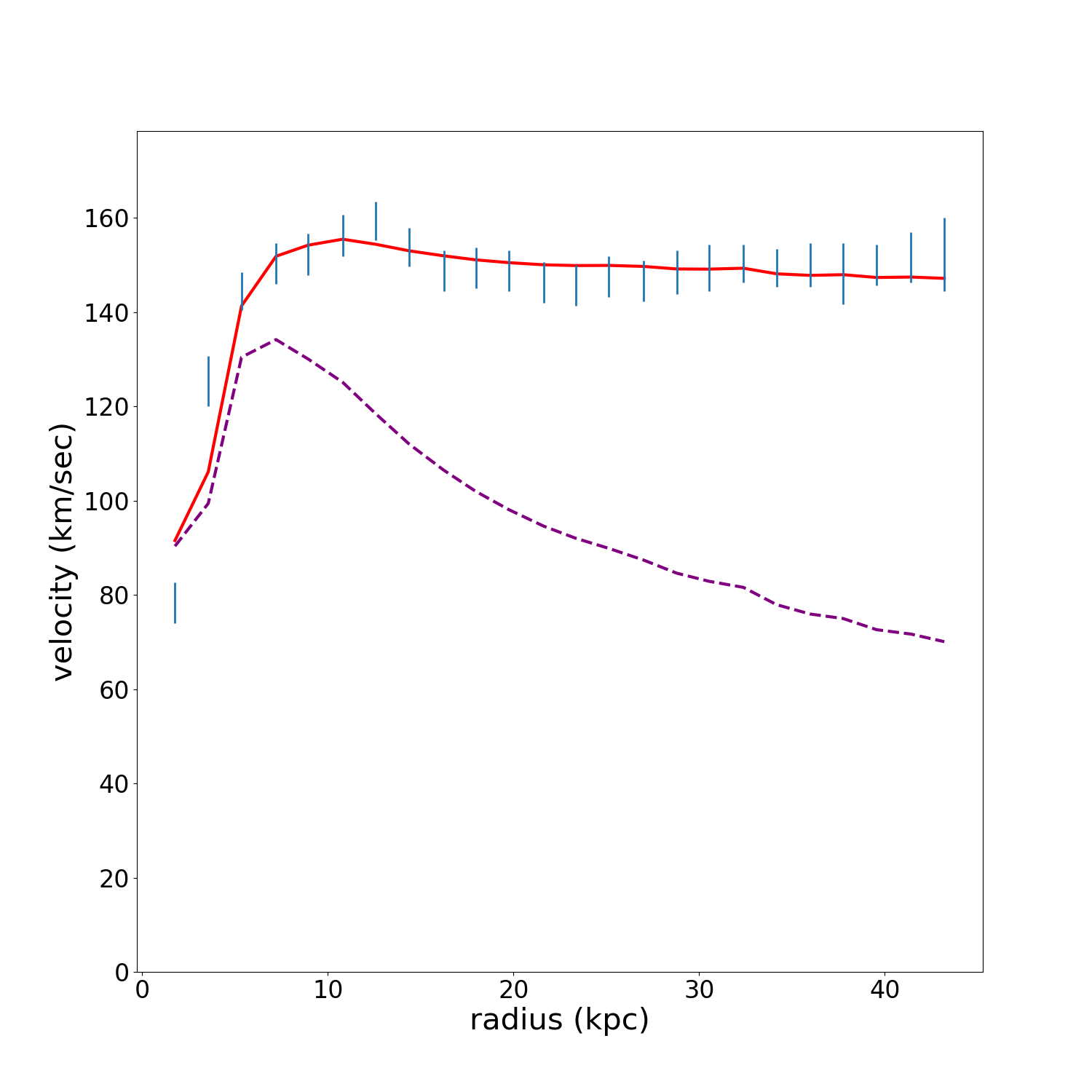}
  \caption{  RC and baryon models from \citep{Maria1}.}
\end{subfigure}
\caption{ RCFM fits  to NGC 3198.
    Lines are as in Fig.~\ref{fig:NGC2403RCFM}. }
\label{fig1super}
\end{figure}

  NGC 3198 is a barred spiral in Ursa Major, which has been well studied. 
  It
  has been considered    a problem galaxy  for MOND, 
  when the distance is a free parameter in fits
\citep{Gent}, as the preferred MOND distance  is 2$\sigma$ different from   that reported from   Cepheids.
However, in the SPARC database the 
RAR fit to  this galaxy at the Cepheid distance reproduces the RC well. This galaxy has no bulge component, as indicated in Table ~\ref{tab:NGC3198}.
  See Fig.~\ref{fig1super} for the RCFM fit. 

 \begin{table} 
        \centering
            \caption{ NGC 3198. Results from fits to the SPARC    data and luminous mass model. \protect\cite{2016Lelli}    \label{tab:NGC3198}} 
        \begin{tabular}{|c|c|c|c|c|c|}
       \hline 
        \hline
        & Model            & $\gamma_d$ & $\gamma_b$   & $\chi^2_r$ & distance (MPC)  \\
        \hline
 \hline
        &Dark Matter Iso   &  0.52   &    --     &  1.31 & 13.7 \\
\hline  
         &RAR	       &  0.77   &	 --     &  2.06 & 13.8 \\
\hline
         &RCFM              &  0.88    &    --     & 1.72  &  13.8 \\
\hline
        \end{tabular}
    \end{table}

    \subsubsection{  NGC 7814 and NGC 891}
 
 The spiral galaxies NGC 7814  and NGC 891
  present an interesting challenge to   dark matter and MOND models. Both galaxies are presented edge-on on the sky, and both have essentially identical RCs, but are  extreme opposites in   morphologies. 
  NGC 7814 is a bulge dominated galaxy and NGC 891 is almost entirely a disk galaxy. 
 \cite{Frat1} asks ``why are these RCs so identical if their dark matter halos are necessarily different to accommodate the differences in the luminous mass?''

\begin{figure} 
\begin{subfigure}{.5\textwidth}
  \centering
  \includegraphics[width=.8\linewidth]{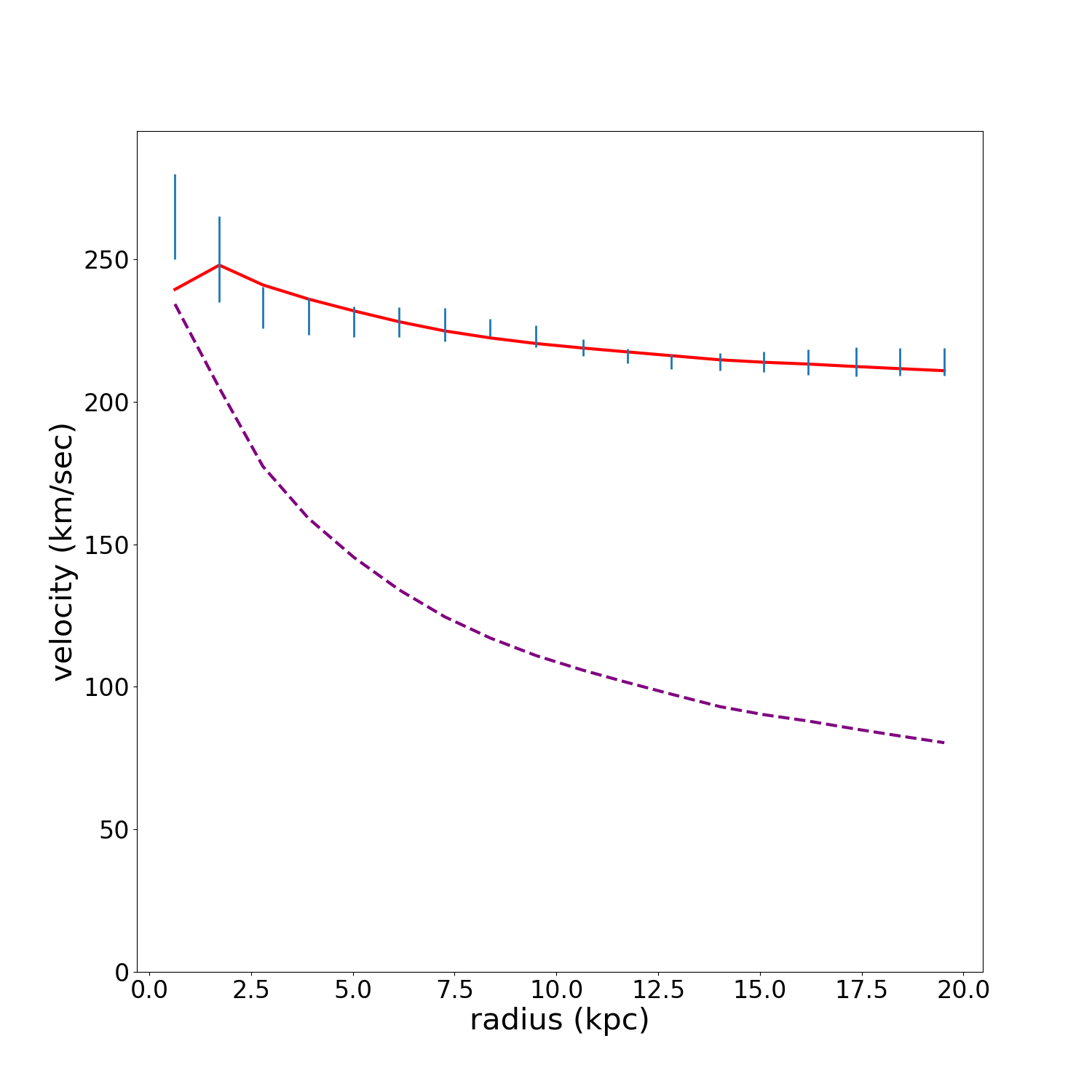}
  \caption{  NGC 7814  }
\end{subfigure}\\
\begin{subfigure}{.5\textwidth}
  \centering
  \includegraphics[width=.8\linewidth]{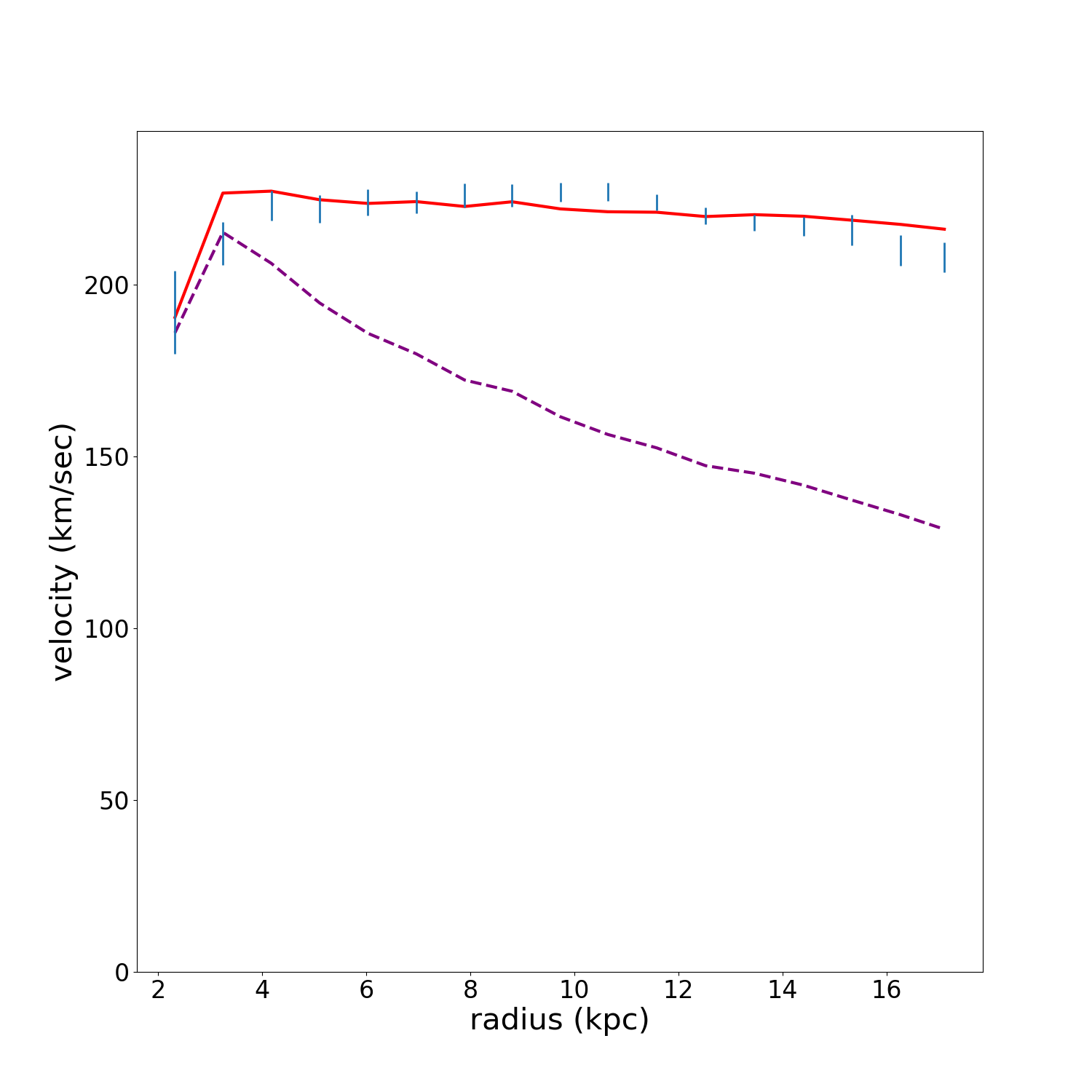}
  \caption{ NGC 891 }
\end{subfigure}
\caption{  Comparison of RCFM fits to galaxies with very similar rotation curves but with very different light distributions, Bulge dominated NGC 7814   and disk dominated NGC 891.   RC and baryon models   from \citep{2016Lelli}.  Lines are as in Fig.~\ref{fig:NGC2403RCFM}.  }  
\label{fitCompare7814_891}
\end{figure}

  In the   RCFM  paradigm the two RCs are  very similar in magnitude,  but   differ  in their inflections  at large radii;  NGC 7814 inflects up, whereas  NGC 891 inflects down.   
  RCFM  
fits to these galaxies are successful, see Fig.\ref{fitCompare7814_891} and Tables~\ref{tab:taleOftwoGalaxies7814} and \ref{tab:taleOftwoGalaxies}. We report here fits to the \citep{Frat1} RC data included in the 
  SPARC database. 
  
  \begin{table} 
        \centering
          \caption{  Bulge dominated NGC 7814. Results from fits to the SPARC    data and luminous mass model. \protect\cite{2016Lelli}    \label{tab:taleOftwoGalaxies7814}} 
        \begin{tabular}{|c|c|c|c|c|}
        \hline 
        \hline
        & Model & $\gamma_d$ & $\gamma_b$& $\chi^2_r$ \\
        \hline
 \hline
        & Dark Matter Iso   &    0.68      &   0.71    & 0.39 \\
\hline
     
         &  RAR	       &   1.17      &	0.52      & 1.334 \\
\hline
         &  RCFM           &   0.38         &	0.68     & 0.63   \\
\hline
        \end{tabular}
    \end{table} 
  
 \begin{table} 
        \centering
            \caption{ Disk dominated NGC 891. Results from fits to the SPARC    data and luminous mass model. \protect\cite{2016Lelli}     \label{tab:taleOftwoGalaxies891}} 
        \begin{tabular}{|c|c|c|c|c|}
       \hline 
        \hline
        & Model             & $\gamma_d$ & $\gamma_b$   & $\chi^2_r$ \\
        \hline
 \hline
        & Dark Matter Iso   &     0.77 &      1.63    & 1.30  \\
\hline
     
         &  RAR	       &    0.32  &	   ---       & 25.16\\
\hline
         &  RCFM           &     0.66  &	---          & 1.89   \\
\hline
        \end{tabular}
        \label{tab:taleOftwoGalaxies}
    \end{table}

\clearpage
\subsubsection{  NGC 5055  }
In this section we compare two different sets of NGC 5055 RC data  from HI,  and their resulting dark matter, RAR  and RCFM fits. 
The first RC   is  from \citep{Batt} as reported in SPARC. Their   luminous model  has     no stellar bulge. 
 The RC data and luminous mass model are reported 
   at  the reliable distance of $9.83 \pm 0.30$ Mpc,  from the   Tip of The Red Giant Branch method.

 \begin{table}
        \centering
          \caption{  Results for a disk dominated NGC 5055. Results from fits to the SPARC    data and luminous mass model. \protect\cite{2016Lelli}, \protect\citet{Batt} .  \label{tab:5055batt}} 
        \begin{tabular}{|c|c|c|c|c|}
        \hline
        \hline
        & Model & $\gamma_d$ & $\gamma_b$& $\chi^2_r$ \\
\hline
         &  Dark matter - Isothermal Halo         &    0.26      &   -     &  6.19  \\
\hline
         &  RAR	                            &    0.56   &   -       & 7.42  \\
\hline  
         &  RCFM                                 &    0.62    &	-     &  6.29 \\
\hline
        \end{tabular}
  \end{table} 

\begin{figure} 
\begin{subfigure}{.5\textwidth}
  \centering
  \includegraphics[width=.8\linewidth]{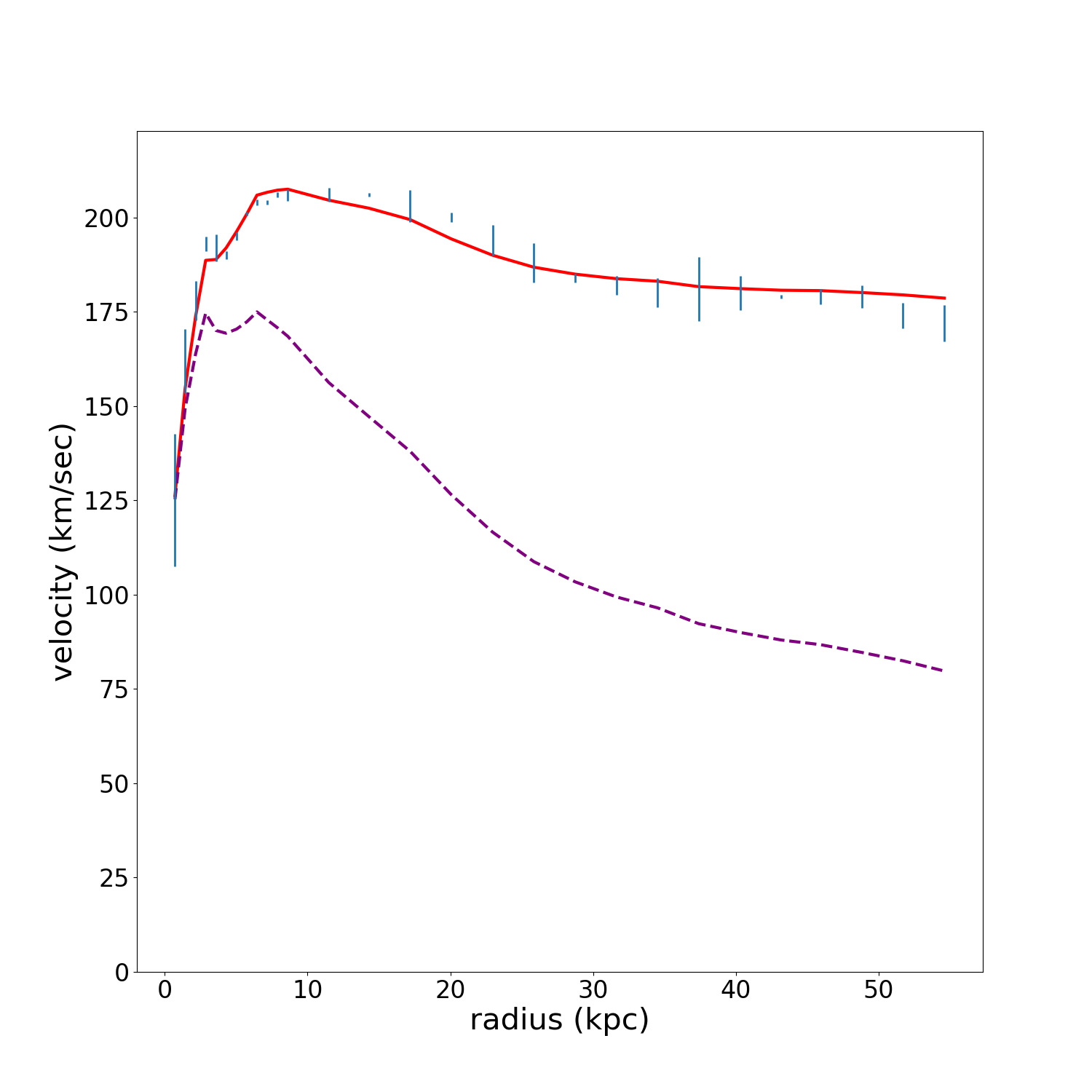}
  \caption{ NGC 5055, RC and baryon model   from  \citep{2016Lelli}.  }
\end{subfigure}\\
\begin{subfigure}{.5\textwidth}
  \centering
  \includegraphics[width=.8\linewidth]{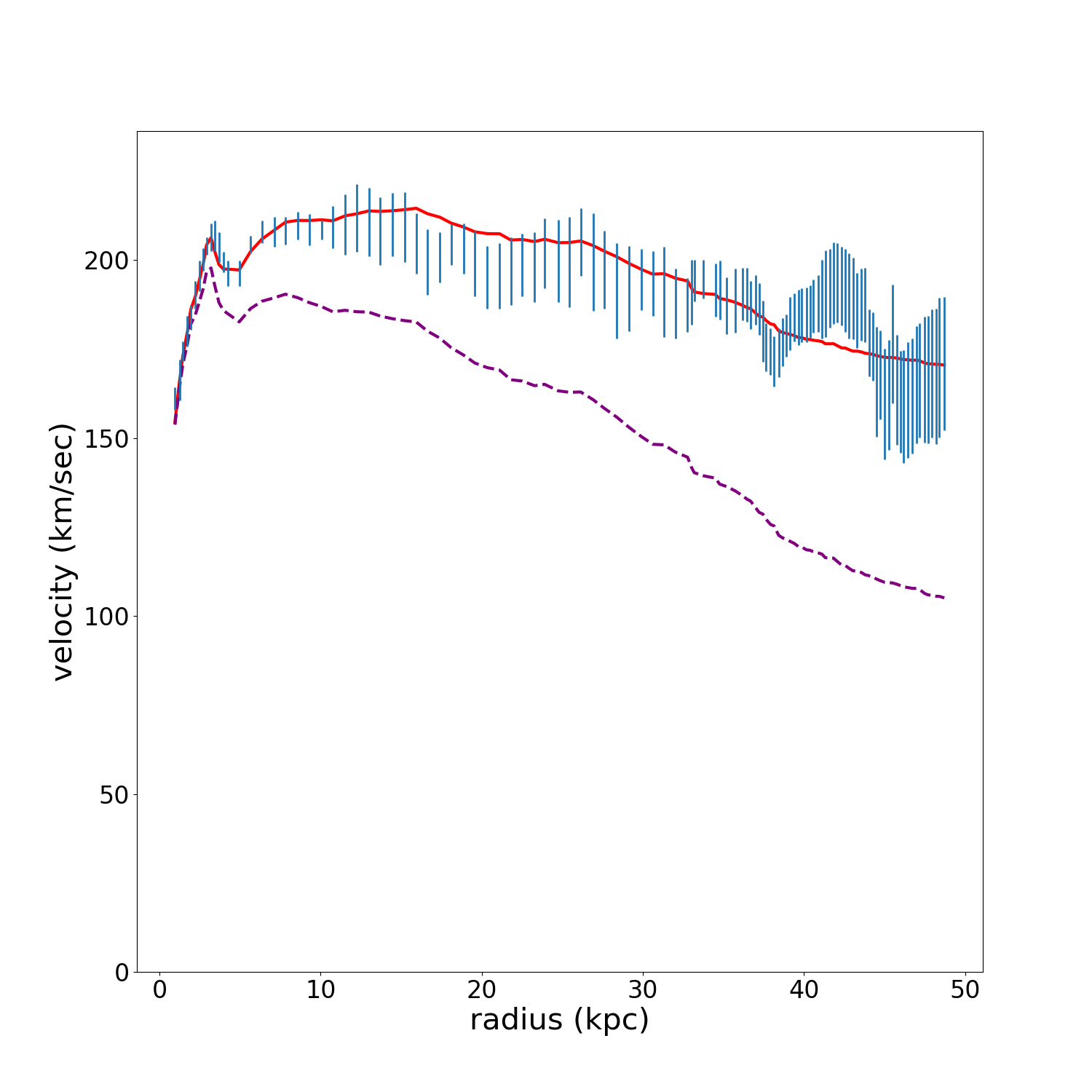}
  \caption{   NGC 5055,  RC and baryon model  from  \citep{Blok1}.  }
\end{subfigure}
\caption{ RCFM comparison fits for  NGC 5055.   Lines are as in Fig.~\ref{fig:NGC2403RCFM}.  }  
\label{fit5055}
\end{figure}

The second RC for     NGC 5055 is from the  THINGS \citep{Blok} at a distance of   $10.1 $ Mpc from the Hubble Flow, including a stellar bulge component.  RCFM fits to both sets of   
  RC data can be seen  in Fig.\ref{fit5055}, at  reported distances, using the reported luminous mass models.   This galaxy serves as an example of  the  under-constrained nature of   galaxy mass modeling   \citep{Conroy}. Details of the fits are given in Table ~\ref{fit5055}. 
  
 \begin{table} 
\centering
          \caption{   Results for a bulge dominated NGC 5055.  RC data and mass model from  \protect\citep{deBlok}.   \label{tab:5055SPARC}} 
 \begin{tabular}{|c|c|c|c|c|}
 \hline
 \hline
        & Model & $\gamma_d$ & $\gamma_b$& $\chi^2_r$ \\
        \hline
 \hline
        & Dark Matter Iso   &    0.79      &   0.11   & 8.13  \\
\hline
     & Dark Matter NFW   &      0.79    &   0.11   & 10.31 \\
\hline
         &  RAR	       &  0.43     &    0.46  &  2.63 \\
\hline
         &  RCFM           &    0.57    &	  0.23   &  1.30  \\
\hline
        \end{tabular} 
  \end{table}

\clearpage

\subsection{ Galaxies for which   model fits fail   \label{Qfactor} }

  Quality flags $(Q)$ are assigned in  the SPARC galaxy database   using the following schema; $Q = 1$ are galaxies with high-quality   rotation curves; $Q = 2$ are
galaxies with minor asymmetries and/or rotation curves of lower
quality; Q = 3 for galaxies with major asymmetries, strong non-circular motions, and/or offsets between HI
and stellar distributions, so that   $Q = 3$ galaxies are not suited to detailed dynamical studies. Of the three models compared in this paper against the 175  SPARC galaxies;    RAR fits all   galaxies,
the RCFM   fits 172 of 175  galaxies (fits fail for  UGC06628 (a $Q=2$ galaxy), and   
F561-1, UGC04305 ($Q=3$ galaxies)), and  the isothermal dark matter model fits  165 of 175 galaxies (fits fail for $Q=2$ galaxies 
 D512-2, UGC00634,    NGC6789, UGC00891, UGC02023, UGC05999, UGC07232 and UGC09992,
 and  $Q=3$ galaxies F567-2 and F574-2).

\subsection{Free Parameter   Correlation  }
\label{FreeCorrel}
 
 To test a  physical interpretation of the    model's free parameter $\alpha$ with respect to  a ratio of   total luminosity $L_{total}$ by the half-light radius $R_{eff}$ (effective radius), we select a 
  subset of  galaxies from the SPARC dataset by the following  criteria: 
 
  \begin{enumerate}
      \item  Select galaxies with the  most accurate and reliable  distance estimates    (tip of the red giant branch and Cepheid variable stars), rejecting all other galaxies. \\
      \item  Select galaxies with inclinations on the sky     in the range   of  $15^o$ to  $80^o$, rejecting galaxies with an inclination  greater than $80^o$ as difficult to assign a true surface brightness profile, and those at inclinations less than $15^o$ as being difficult to  report line of sight Doppler shifts accurately.\\
      \item   Exclude galaxies  with quality factor $Q=3$ as not suited to dynamical studies due to     asymmetries,  non-circular motions, and/or offsets between stars and gas, using assigned   SPARC $Q$ factors as in    \citep{2016Lelli}.
  \end{enumerate}

\begin{figure}
\begin{subfigure}{.5\textwidth}
  \centering
  \includegraphics[width=.99\linewidth]{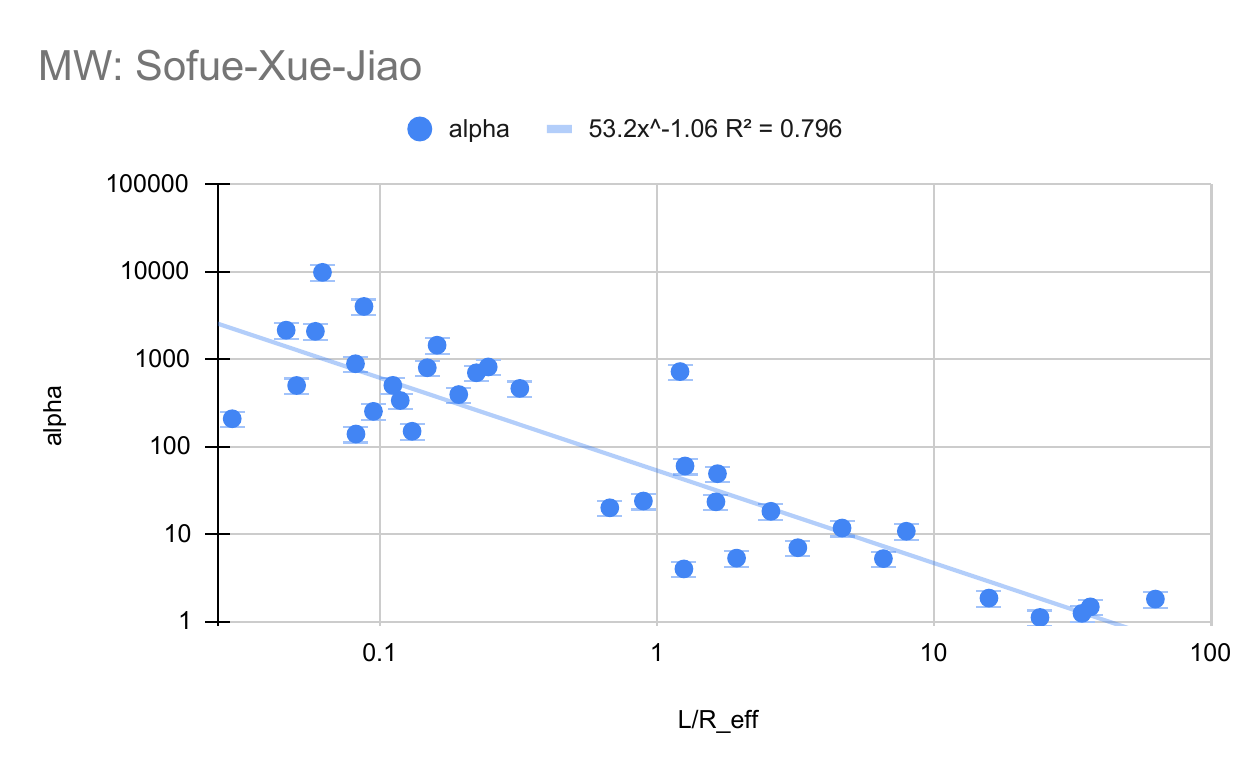}
\end{subfigure}\\
\begin{subfigure}{.5\textwidth}
  \centering
  \includegraphics[width=.99\linewidth]{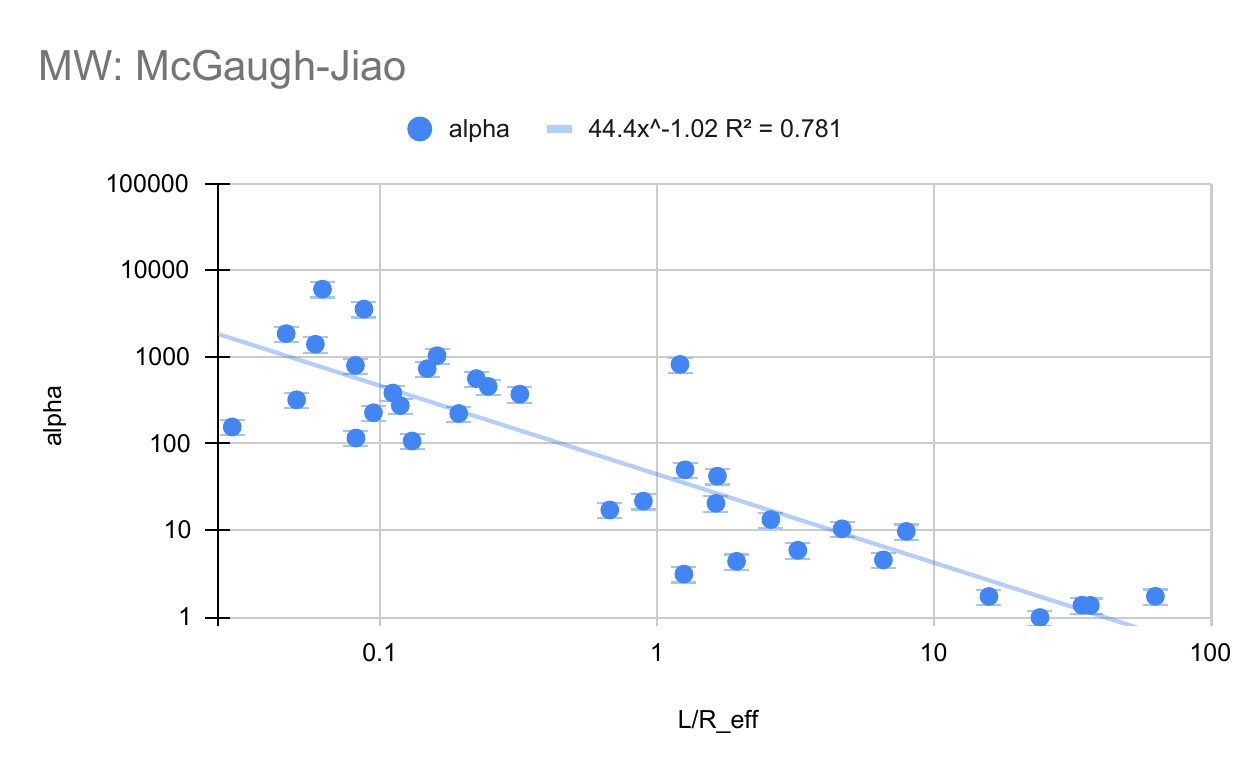}
\end{subfigure}
\caption{ The RCFM  results for the free parameter ($\alpha$)  vs. a ratio of photometric quantities ($L/R_{eff}$) (total luminosity by effective or half-light radius) compared for two different choices of Milky Way baryon model (see Table~\ref{MW_dats}). These images represent fits on  a subset of  SPARC Galaxies, selected for based on best distance estimates. The 36 selected galaxies are listed in     Table~\ref{tab:lobes}.  \label{alpha2}}
\end{figure}

    By this process,   a training set of  $36$ galaxies are selected, as reported in Table~\ref{tab:lobes}.   We then plot the subset's  $\alpha$   values versus the ratio of   $L_{total}/R_{eff}$, and fit the distribution with a power law as shown in Fig.~\ref{alpha2}. 
 The  total luminosity    and  effective radii numbers used are    as reported in the SPARC database. Luminosity measurements are taken in the 
  wavelength of $3.6 \mu m$,  assuming a solar
absolute magnitude of 3.24 at $3.6 \mu m$ \citep{oh2008high}   in units of $10^9$ solar  $L_\odot$. As can be seen in in Fig.~\ref{alpha2} the subset appears to be highly correlated to the ratio of photometric quantities, though slightly different functionally for the choice of Milky Way.

  \begin{table}
     \centering
        \begin{tabular}{|c|c|c|c|c|}
        \hline 
        & Model & $\gamma_d$ & $\gamma_b$& $\chi^2_{red}$ \\
        \hline
         &   RAR & 0.46 &0.74  & 3.89 \\
         \hline
         &  RCFM  (MW: Sofue-Xue-Jiao)     &  1.04 &	0.87  & 1.60 \\
           \hline
            &  RCFM  (MW: McGaugh-Jiao)     &  1.00 &	0.87  & 1.65 \\
           \hline
            &  Dark Matter &  0.50  &0.66 & 1.59 \\
           \hline
        \end{tabular}
         \caption{{\bf Average Fit Results Galaxy Subset}  \\
         A subset of $36$ galaxies selected based on best distance estimates and highest quality rotation curve data.   Reduced $\chi^2$ values  do not reflect    galaxies which  fail to  converge on a fit; for which,  the RCFM and RAR have none, and the isothermal dark matter model   has two  (NGC6789, UGC07232).   The average stellar mass-to-light ratios $\gamma_i$ are presented in units of  solar mass per solar luminosity $M_\odot/L_\odot$. The RCFM spherical simplification used in RCFM fits for computational and formulaic simplicity, can be seen to yield a disk mass-to-light ratio   $\gamma_d$ which is  approximately a factor of $2$ larger than those from RAR and Dark Matter fits. This is a well known  artifact of the spherical assumption, and could be removed with a full disk geometry at the cost of additional computations and free-parameters, see Sec.~\ref{GeomSphere} for more information. } 
        \label{tab:lobes}
    \end{table}

 \begin{table*}
 \caption{Subset of galaxies selected  from   SPARC sample, for testing the RCFM free parameter $\alpha$ as seen in Figures~\ref{tab:lobes}}
\begin{tabular}{|l|l|l|l|l|l|}
\hline
  &     MW:   &   MW:     &    &     &     \\  
Galaxy  &        McGaugh-Jiao &     Sofue-Xue-Jiao   & Distance   & Distance   & Inc   \\  
 Name & $\chi^2_{r}$  & $\chi^2_{r}$  &   (Mpc) &   Method \footnote{ Distance method:  2 = tip of the red giant branch, 3 = Cepheids. Distance and inclination   information from \citep{2016Lelli}.} &   (deg)\\ \hline
CamB        & 0.52                                 & 0.36                                   & 3.36           & 2                   & 65        \\ \hline
D564-8      & 0.03                                 & 0.07                                   & 8.79           & 2                   & 63        \\ \hline
D631-7      & 0.52                                 & 0.36                                   & 7.72           & 2                   & 59        \\ \hline
DDO154      & 8.65                                 & 8.66                                   & 4.04           & 2                   & 64        \\ \hline
DDO168      & 4.17                                 & 4.12                                   & 4.25           & 2                   & 63        \\ \hline
ESO444-G084 & 1.23                                 & 0.63                                   & 4.83           & 2                   & 32        \\ \hline
IC2574      & 1.88                                 & 2.05                                   & 3.91           & 2                   & 75        \\ \hline
NGC0024     & 0.82                                 & 0.80                                   & 7.3            & 2                   & 64        \\ \hline
NGC0055     & 2.24                                 & 2.40                                   & 2.11           & 2                   & 77        \\ \hline
NGC0247     & 2.26                                 & 2.24                                   & 3.7            & 2                   & 74        \\ \hline
NGC0300     & 0.41                                 & 0.41                                   & 2.08           & 2                   & 42        \\ \hline
NGC2403     & 12.05                                & 11.52                                  & 3.16           & 2                   & 63        \\ \hline
NGC2683     & 0.95                                 & 0.97                                   & 9.81           & 2                   & 80        \\ \hline
NGC2841     & 1.51                                 & 1.35                                   & 14.1           & 3                   & 76        \\ \hline
NGC2915     & 0.62                                 & 0.58                                   & 4.06           & 2                   & 56        \\ \hline
NGC2976     & 0.45                                 & 0.46                                   & 3.58           & 2                   & 61        \\ \hline
NGC3109     & 0.30                                 & 0.27                                   & 1.33           & 2                   & 70        \\ \hline
NGC3198     & 1.42                                 & 1.50                                   & 13.8           & 3                   & 73        \\ \hline
NGC3741     & 0.76                                 & 0.63                                   & 3.21           & 2                   & 70        \\ \hline
NGC4214     & 1.55                                 & 1.32                                   & 2.87           & 2                   & 15        \\ \hline
NGC5055     & 6.29                                 & 6.10                                   & 9.9            & 2                   & 55        \\ \hline
NGC6503     & 1.14                                 & 1.14                                   & 6.26           & 2                   & 74        \\ \hline
NGC6789     & 0.86                                 & 0.94                                   & 3.52           & 2                   & 43        \\ \hline
NGC6946     & 2.08                                 & 1.98                                   & 5.52           & 2                   & 38        \\ \hline
NGC7331     & 1.08                                 & 1.09                                   & 14.7           & 3                   & 75        \\ \hline
NGC7793     & 0.72                                 & 0.72                                   & 3.61           & 2                   & 47        \\ \hline
UGC04483    & 0.38                                 & 0.44                                   & 3.34           & 2                   & 58        \\ \hline
UGC07232    & 1.07                                 & 1.02                                   & 2.83           & 2                   & 59        \\ \hline
UGC07524    & 1.40                                 & 1.41                                   & 4.74           & 2                   & 46        \\ \hline
UGC07559    & 0.17                                 & 0.20                                   & 4.97           & 2                   & 61        \\ \hline
UGC07577    & 0.06                                 & 0.06                                   & 2.59           & 2                   & 63        \\ \hline
UGC07866    & 0.08                                 & 0.07                                   & 4.57           & 2                   & 44        \\ \hline
UGC08490    & 0.15                                 & 0.13                                   & 4.65           & 2                   & 50        \\ \hline
UGC08837    & 0.62                                 & 0.65                                   & 7.21           & 2                   & 80        \\ \hline
UGCA442     & 0.74                                 & 0.73                                   & 4.35           & 2                   & 64        \\ \hline
UGCA444     & 0.10                                 & 0.10                                   & 0.98           & 2                   & 78        \\ \hline
\label{TSet}
\end{tabular}
\end{table*}
\newpage

 \section{  Conclusions }
  \label{sec:conclu} 
 
    At this time, there are many discrepancies between   cosmology theory and observations
 which are resolved by 
 dark matter models \citep{2010dmp..book.....S,Tully:2014gfa,Naidu_2022}.  
    In
 this paper we  address only one such discrepancy,  the flat-rotation curve problem of spiral galaxies.
 This choice is due to the   clear symmetry presented in the rotation curve data. 
 Other dark matter problems require cosmological model assumptions, beyond      the scope of the current paper. 
 
Here we reinterpret  the flat-rotation curve problem not as a problem of missing mass but rather,  at the level of the data, as  a problem  of  relative frame effects due to  our home galaxy. We  predict that the flat-rotation curve problem is an artifact of misinterpretation of Doppler shifted spectra from external galaxies, removing the need for dark matter halos on  galaxies. 
    This framing places primary importance on  the role of the Milky Way,   and compactly explains both why the  Milky Way sits roughly at the  inflection  point in  the Universal Rotation Curve spectrum of 1,100 galaxies \citep{salucci} and why MOND is both successful and limited.

  The rotation curve fitting model (RCFM) presented here   reproduces the fitting successes of MOND, RAR  (Radial Acceleration Relation)  and dark matter models on a  sample of   175  well studied galaxy rotation curves \citep{2016Lelli}, but does not   modify classical  physics. 
  Independent of the interpretation of the various quantities appearing in the formulae presented in this paper, this is a one parameter fit to the data.
  What is more, the RCFM  free parameter is highly correlated with a ratio of   observable photometric  parameters, which makes this the first direct constraint to luminous mass modeling of galaxies from Doppler shifted spectra.
  
   We emphasize that the only input to the RCFM is the luminous mass. 
    In contrast to previous investigations of gravitational redshifts in this context, which  used Galilean subtraction \citep{MTW}, this approach sets the frame comparison from the center of the galaxies, using  Lorentz-type boosts between galaxy frames in radius. 
    In this way, using gravitational redshifts to define frames, we can    explain the discrepancy between the ``flat-rotation'' curves and the Keplerian luminous mass predictions  with only one free parameter.  
  Our results are presented in this paper in comparison to dark matter and RAR fit   results for the same galaxy data, from the SPARC database of 175  rotation curves \citep{2016Lelli}. RAR has a simpler functional form than MOND, though the same physics paradigm.
 The  recent \emph{Gaia} DR3 data release   \cite{jiao2023detection}      has shown  that   the rotation curve of the Milky Way demonstrates  a Keplerian decline  from  $10$ - $26.5$ kpc,   consistent with the RCFM  model paradigm presented here.    The upcoming Large  Survey of Space and Time  at the   Vera C. Rubin Observatory \citep{Ivezić_2019} can falsify this model if  the Milky Way rotation curve beyond $26.5$ kpc is found to require a dark matter halo.

 \section[]{Acknowledgments} 
This work is dedicated to Emmett Till. 
 We acknowledge and express     gratitude to the first nations peoples on whose ceded and unceded lands this was written; including but not limited  to  the 
 the Coast Salish bands of the Puget Sound, 
 the Cheyenne and Arapaho Tribes and  Ute  Tribes of Colorado,   the Algonquian and Iroquoian  Peoples of Massachusetts and New York, and the Navajo and Pueblo Nations of New Mexico. 
  The authors would like to thank    V.\,P.\,  Nair, T.\, Boyer,  M.\, Kaku, I.\, Chavel,  R.\, Walterbos, N.\,P.\, Vogt,
  S.\, McGaugh, A.\, Klypin, T.\, Quinn, S.\ Tuttle, M.\, Juric,  R.\, Rivera, N. Oblath,  J.\, Formaggio, J.\, Conrad,  P.\, Fisher, Y.\, Sofue, C.\, Mihos, and M.\, Merrifield. 
  \clearpage

\clearpage
\onecolumn
 \begin{landscape}
\begin{center} 
\begin{longtable}{|l|l|l|l|l|l|l|l|l|l|l|}
\caption{ Fit Results for  the SPARC galaxies Rotation Curve} \label{table:M2Light}  \\

\hline \multicolumn{1}{|c|}{\textbf{\textbf{MW:}SXJ}} & \multicolumn{3}{c|}{\textbf{Dark Matter}} & \multicolumn{3}{c|}{\textbf{RCFM}} & \multicolumn{3}{c|}{\textbf{RAR}} & \multicolumn{1}{|c|}{} \\ \hline 
 \multicolumn{1}{|c|}{\textbf{Galaxy}} & \multicolumn{1}{c|}{$\chi^2_{r}$ } & \multicolumn{1}{c|}{$\gamma_{disk}$ } & \multicolumn{1}{c|}{$\gamma_{bulge}$ } 
 & \multicolumn{1}{c|}{$\chi^2_{r}$ } & \multicolumn{1}{c|}{$\gamma_{disk}$ } & \multicolumn{1}{c|}{$\gamma_{bulge}$ }
& \multicolumn{1}{c|}{$\chi^2_{r}$ } & \multicolumn{1}{c|}{$\gamma_{disk}$ } & \multicolumn{1}{c|}{ $\gamma_{bulge}$ } & {\textbf{Q}} \\ \hline             
\endfirsthead

\multicolumn{3}{c}%
{{\bfseries \tablename\ \thetable{} -- continued from previous page}} \\
\hline \multicolumn{1}{|c|}{\textbf{\textbf{MW:}SXJ}} & \multicolumn{3}{c|}{\textbf{Dark Matter}} & \multicolumn{3}{c|}{\textbf{RCFM}} & \multicolumn{3}{c|}{\textbf{RAR}} & \multicolumn{1}{|c|}{} \\ \hline 
 \multicolumn{1}{|c|}{\textbf{Galaxy}} & \multicolumn{1}{c|}{$\chi^2_{r}$ } & \multicolumn{1}{c|}{$\gamma_{disk}$ } & \multicolumn{1}{c|}{$\gamma_{bulge}$ } 
 & \multicolumn{1}{c|}{$\chi^2_{r}$ } & \multicolumn{1}{c|}{$\gamma_{disk}$ } & \multicolumn{1}{c|}{$\gamma_{bulge}$ }
& \multicolumn{1}{c|}{$\chi^2_{r}$ } & \multicolumn{1}{c|}{$\gamma_{disk}$ } & \multicolumn{1}{c|}{ $\gamma_{bulge}$ } & {\textbf{Q}} \\ \hline  
\endhead

\hline \multicolumn{11}{|r|}{{Continued on next page}} \\ \hline
\endfoot

\hline \hline
\endlastfoot

 Averages:  & 1.90           & 0.51            & 0.11             & 2.39           & 1.13            & 0.78             & 4.22      & 0.64            & 0.73             & -- \\ \hline
CamB             & 2.99           & 0.34            & 0.00             & 0.23           & 7.91E-05            & …                & 5.76      & 0.34            & …                & 2.00             \\ \hline
D512-2           & NAN            &…               & …               & 0.21           & 1.48            & …                & 0.37      & 0.48            & …                & 2.00             \\ \hline
D564-8           & 0.26           & 0.49            & 0.00             & 0.11           & 1.21            & …                & 3.16      & 0.40            & …                & 2.00             \\ \hline
D631-7           & 0.85           & 0.40            & 0.00             & 0.30           & 0.26            & …                & 15.87     & 0.20            & …                & 1.00             \\ \hline
DDO064           & 0.48           & 0.51            & 0.00             & 0.45           & 1.58            & …                & 0.33      & 0.48            & …                & 1.00             \\ \hline
DDO154           & 3.23           & 0.41            & 0.00             & 10.58          & 1.18            & …                & 3.48      & 0.19            & …                & 2.00             \\ \hline
DDO161           & 0.36           & 0.50            & 0.00             & 0.62           & 0.97            & …                & 1.47      & 0.23            & …                & 1.00             \\ \hline
DDO168           & 6.03           & 0.45            & 0.00             & 4.36           & 0.76            & …                & 19.71     & 0.46            & …                & 2.00             \\ \hline
DDO170           & 2.45           & 0.49            & 0.00             & 3.19           & 1.84            & …                & 4.92      & 0.79            & …                & 2.00             \\ \hline
ESO079-G014      & 1.61           & 0.51            & 0.00             & 3.86           & 1.09            & …                & 4.33      & 0.50            & …                & 1.00             \\ \hline
ESO116-G012      & 1.27           & 0.55            & 0.00             & 1.03           & 1.04            & …                & 2.44      & 0.35            & …                & 1.00             \\ \hline
ESO444-G084      & 2.74           & 0.51            & 0.00             & 0.40           & 1.86            & …                & 3.25      & 0.42            & …                & 2.00             \\ \hline
ESO563-G021      & 12.37          & 0.55            & 0.00             & 16.41          & 1.00            & …                & 28.84     & 0.43            & …                & 1.00             \\ \hline
F561-1           & 1.14           & 0.50            & 0.00             & NAN            &  …                & …                & 1.56      & 0.52            & …                & 3.00             \\ \hline
F563-1           & 0.70           & 0.50            & 0.00             & 0.95           & 2.06            & …                & 1.50      & 0.56            & …                & 1.00             \\ \hline
F563-V1          & 0.98           & 0.50            & 0.00             & 0.29           & 0.99            & …                & 0.88      & 0.48            & …                & 3.00             \\ \hline
F563-V2          & 0.47           & 0.51            & 0.00             & 0.11           & 2.20            & …                & 0.99      & 0.59            & …                & 1.00             \\ \hline
F565-V2          & 0.16           & 0.50            & 0.00             & 0.31           & 2.22            & …                & 0.47      & 0.50            & …                & 2.00             \\ \hline
F567-2           & NAN            &…              & …               & 0.50           & 1.31            & …                & 2.20      & 0.56            & …                & 3.00             \\ \hline
F568-1           & 0.18           & 0.50            & 0.00             & 0.72           & 1.91            & …                & 1.29      & 0.61            & …                & 1.00             \\ \hline
F568-3           & 1.18           & 0.47            & 0.00             & 1.79           & 1.31            & …                & 3.06      & 0.41            & …                & 1.00             \\ \hline
F568-V1          & 0.19           & 0.51            & 0.00             & 0.14           & 2.14            & …                & 1.04      & 0.81            & …                & 1.00             \\ \hline
F571-8           & 0.75           & 0.37            & 0.00             & 2.04           & 0.17            & …                & 41.61     & 0.11            & …                & 1.00             \\ \hline
F571-V1          & 0.18           & 0.52            & 0.00             & 0.20           & 1.49            & …                & 0.29      & 0.50            & …                & 2.00             \\ \hline
F574-1           & 0.35           & 0.48            & 0.00             & 1.42           & 1.54            & …                & 2.50      & 0.71            & …                & 1.00             \\ \hline
F574-2           & NAN            &…                & …              & 0.19           & 0.75            & …                & 0.09      & 0.49            & …                & 3.00             \\ \hline
F579-V1          & 0.06           & 0.51            & 0.00             & 1.07           & 1.63            & …                & 2.56      & 0.63            & …                & 1.00             \\ \hline
F583-1           & 0.36           & 0.50            & 0.00             & 1.04           & 1.87            & …                & 2.66      & 0.91            & …                & 1.00             \\ \hline
F583-4           & 0.42           & 0.51            & 0.00             & 0.28           & 1.30            & …                & 0.13      & 0.48            & …                & 1.00             \\ \hline
IC2574           & 2.51           & 0.77            & 0.00             & 2.27           & 1.10            & …                & 1.44      & 0.07            & …                & 2.00             \\ \hline
IC4202           & 8.67           & 0.50            & 0.31             & 32.72          & 1.09            & 0.74             & 41.91     & 1.60            & 0.34             & 1.00             \\ \hline
KK98-251         & 0.35           & 0.49            & 0.00             & 0.42           & 1.67            & …                & 1.23      & 0.44            & …                & 2.00             \\ \hline
NGC0024          & 0.37           & 0.54            & 0.00             & 0.73           & 1.39            & …                & 0.85      & 1.01            & …                & 1.00             \\ \hline
NGC0055          & 0.55           & 0.39            & 0.00             & 2.86           & 1.01            & …                & 1.58      & 0.19            & …                & 2.00             \\ \hline
NGC0100          & 0.10           & 0.49            & 0.00             & 0.10           & 0.93            & …                & 1.29      & 0.28            & …                & 1.00             \\ \hline
NGC0247          & 2.38           & 0.46            & 0.00             & 2.18           & 1.53            & …                & 3.06      & 0.78            & …                & 2.00             \\ \hline
NGC0289          & 1.97           & 0.45            & 0.00             & 1.78           & 0.74            & …                & 2.13      & 0.92            & …                & 2.00             \\ \hline
NGC0300          & 0.47           & 0.50            & 0.00             & 0.42           & 1.14            & …                & 0.91      & 0.40            & …                & 2.00             \\ \hline
NGC0801          & 7.07           & 0.64            & 0.00             & 7.38           & 0.77            & …                & 7.75      & 1.33            & …                & 1.00             \\ \hline
NGC0891          & 4.24           & 0.27            & 0.60             & 1.89           & 0.66            & 3.77E-05             & 7.37      & 0.33            & 0.40             & 1.00             \\ \hline
NGC1003          & 2.67           & 0.77            & 0.00             & 3.42           & 0.77            & …                & 4.67      & 0.37            & …                & 1.00             \\ \hline
NGC1090          & 1.54           & 0.53            & 0.00             & 2.27           & 0.81            & …                & 2.78      & 0.74            & …                & 1.00             \\ \hline
NGC1705          & 0.08           & 0.49            & 0.00             & 0.13           & 1.25            & …                & 0.37      & 1.22            & …                & 3.00             \\ \hline
NGC2366          & 1.09           & 0.39            & 0.00             & 2.34           & 1.06            & …                & 1.93      & 0.24            & …                & 3.00             \\ \hline
NGC2403          & 10.94          & 0.83            & 0.00             & 10.73          & 0.86            & …                & 14.14     & 0.51            & …                & 1.00             \\ \hline
NGC2683          & 1.89           & 0.60            & 0.70             & 1.01           & 0.88            & 0.44             & 3.37      & 0.55            & 0.73             & 2.00             \\ \hline
NGC2841          & 1.58           & 0.60            & 0.66             & 1.36           & 0.91            & 1.10             & 1.52      & 0.81            & 0.93             & 1.00             \\ \hline
NGC2903          & 6.67           & 0.43            & 0.00             & 7.45           & 0.62            & …                & 20.64     & 0.21            & …                & 1.00             \\ \hline
NGC2915          & 0.71           & 0.41            & 0.00             & 0.63           & 0.56            & …                & 4.02      & 0.32            & …                & 2.00             \\ \hline
NGC2955          & 4.21           & 0.36            & 0.91             & 4.45           & 0.40            & 0.88             & 3.91      & 0.37            & 0.84             & 1.00             \\ \hline
NGC2976          & 0.33           & 0.48            & 0.00             & 0.46           & 0.91            & …                & 1.73      & 0.35            & …                & 2.00             \\ \hline
NGC2998          & 1.47           & 0.36            & 0.00             & 3.74           & 0.87            & …                & 2.94      & 0.82            & …                & 1.00             \\ \hline
NGC3109          & 0.23           & 0.50            & 0.00             & 0.28           & 2.05            & …                & 4.13      & 0.21            & …                & 1.00             \\ \hline
NGC3198          & 1.31           & 0.52            & 0.00             & 1.72           & 0.88            & …                & 2.06      & 0.77            & …                & 1.00             \\ \hline
NGC3521          & 0.20           & 0.52            & 0.00             & 0.77           & 0.71            & …                & 0.51      & 0.46            & …                & 1.00             \\ \hline
NGC3726          & 2.54           & 0.61            & 0.00             & 2.41           & 0.76            & …                & 2.98      & 0.47            & …                & 2.00             \\ \hline
NGC3741          & 0.90           & 0.71            & 0.00             & 0.65           & 0.71            & …                & 0.77      & 0.31            & …                & 1.00             \\ \hline
NGC3769          & 0.88           & 0.43            & 0.00             & 0.71           & 0.71            & …                & 0.95      & 0.41            & …                & 2.00             \\ \hline
NGC3877          & 3.47           & 0.31            & 0.00             & 9.07           & 0.87            & …                & 10.22     & 0.40            & …                & 2.00             \\ \hline
NGC3893          & 0.69           & 0.46            & 0.00             & 0.57           & 0.73            & …                & 1.00      & 0.45            & …                & 1.00             \\ \hline
NGC3917          & 1.64           & 0.53            & 0.00             & 2.70           & 1.14            & …                & 4.60      & 0.55            & …                & 1.00             \\ \hline
NGC3949          & 0.84           & 0.52            & 0.00             & 0.76           & 0.73            & …                & 0.55      & 0.44            & …                & 2.00             \\ \hline
NGC3953          & 0.56           & 0.49            & 0.00             & 0.76           & 0.90            & …                & 3.42      & 0.59            & …                & 1.00             \\ \hline
NGC3972          & 0.80           & 0.45            & 0.00             & 2.08           & 1.03            & …                & 2.07      & 0.50            & …                & 1.00             \\ \hline
NGC3992          & 0.71           & 0.55            & 0.00             & 1.68           & 1.04            & …                & 3.47      & 0.76            & …                & 1.00             \\ \hline
NGC4010          & 2.16           & 0.46            & 0.00             & 1.90           & 0.85            & …                & 2.74      & 0.36            & …                & 2.00             \\ \hline
NGC4013          & 0.84           & 0.58            & 0.83             & 1.53           & 0.54            & 1.43             & 1.81      & 0.35            & 0.79             & 2.00             \\ \hline
NGC4051          & 1.73           & 0.48            & 0.00             & 1.91           & 0.81            & …                & 2.49      & 0.45            & …                & 2.00             \\ \hline
NGC4068          & 0.82           & 0.49            & 0.00             & 0.27           & 0.80            & …                & 2.52      & 0.38            & …                & 2.00             \\ \hline
NGC4085          & 5.28           & 0.36            & 0.00             & 2.83           & 0.58            & …                & 9.09      & 0.35            & …                & 2.00             \\ \hline
NGC4088          & 0.64           & 0.46            & 0.00             & 0.87           & 0.63            & …                & 0.66      & 0.40            & …                & 1.00             \\ \hline
NGC4100          & 1.21           & 0.57            & 0.00             & 1.90           & 0.95            & …                & 1.66      & 0.76            & …                & 1.00             \\ \hline
NGC4138          & 5.52           & 0.55            & 0.68             & 0.75           & 0.98            & 0.00             & 2.49      & 0.55            & 0.69             & 2.00             \\ \hline
NGC4157          & 0.45           & 0.53            & 0.66             & 0.61           & 0.69            & 0.56             & 0.72      & 0.43            & 0.64             & 1.00             \\ \hline
NGC4183          & 0.14           & 0.53            & 0.00             & 0.53           & 1.24            & …                & 1.13      & 0.79            & …                & 1.00             \\ \hline
NGC4214          & 0.77           & 0.50            & 0.00             & 1.19           & 1.01            & …                & 1.06      & 0.46            & …                & 2.00             \\ \hline
NGC4217          & 2.23           & 0.60            & 0.30             & 1.33           & 1.13            & 0.46             & 3.17      & 1.17            & 0.17             & 1.00             \\ \hline
NGC4389          & 6.30           & 0.29            & 0.00             & 0.20           & 0.31            & …                & 9.31      & 0.30            & …                & 3.00             \\ \hline
NGC4559          & 0.35           & 0.50            & 0.00             & 0.34           & 0.77            & …                & 0.50      & 0.52            & …                & 1.00             \\ \hline
NGC5005          & 0.10           & 0.49            & 0.60             & 0.07           & 0.62            & 0.69             & 0.09      & 0.54            & 0.56             & 1.00             \\ \hline
NGC5033          & 7.15           & 0.52            & 0.36             & 6.39           & 0.85            & 0.52             & 8.02      & 1.03            & 0.43             & 1.00             \\ \hline
NGC5055          & 6.19           & 0.26            & 0.00             & 6.29           & 0.62            & …                & 7.42      & 0.56            & …                & 1.00             \\ \hline
NGC5371          & 1.96           & 0.42            & 0.00             & 12.55          & 0.78            & …                & 10.16     & 3.30            & …                & 1.00             \\ \hline
NGC5585          & 5.18           & 0.55            & 0.00             & 6.35           & 0.75            & …                & 6.82      & 0.22            & …                & 1.00             \\ \hline
NGC5907          & 3.30           & 0.42            & 0.00             & 7.88           & 0.94            & …                & 7.73      & 1.08            & …                & 1.00             \\ \hline
NGC5985          & 2.54           & 0.53            & 0.64             & 5.91           & 1.01            & 2.12             & 6.97      & 0.63            & 3.32             & 1.00             \\ \hline
NGC6015          & 9.28           & 0.64            & 0.00             & 13.06          & 0.99            & …                & 10.87     & 1.12            & …                & 2.00             \\ \hline
NGC6195          & 1.81           & 0.41            & 0.79             & 2.95           & 0.41            & 0.81             & 2.26      & 0.32            & 0.85             & 1.00             \\ \hline
NGC6503          & 1.29           & 0.34            & 0.00             & 1.44           & 0.75            & …                & 2.98      & 0.45            & …                & 1.00             \\ \hline
NGC6674          & 1.47           & 0.87            & 0.74             & 3.98           & 0.70            & 1.87             & 10.64     & 0.95            & 1.30             & 1.00             \\ \hline
NGC6789          & NAN            &…              &…                & 0.57           & 1.41            & …                & 5.90      & 0.60            & …                & 2.00             \\ \hline
NGC6946          & 1.55           & 0.57            & 0.61             & 1.75           & 0.64            & 0.58             & 1.53      & 0.64            & 0.71             & 1.00             \\ \hline
NGC7331          & 0.95           & 0.45            & 0.65             & 1.05           & 0.53            & 1.17             & 1.29      & 0.32            & 0.60             & 1.00             \\ \hline
NGC7793          & 0.79           & 0.58            & 0.00             & 0.75           & 0.89            & …                & 1.01      & 0.55            & …                & 1.00             \\ \hline
NGC7814          & 0.46           & 0.52            & 0.60             & 0.63           & 0.38            & 0.68             & 1.33      & 1.17            & 0.52             & 1.00             \\ \hline
PGC51017         & 5.98           & 0.41            & 0.00             & 2.31           & 0.80            & …                & 4.57      & 0.44            & …                & 3.00             \\ \hline
UGC00128         & 3.56           & 0.57            & 0.00             & 6.16           & 1.64            & …                & 6.25      & 2.49            & …                & 1.00             \\ \hline
UGC00191         & 1.97           & 0.49            & 0.00             & 2.45           & 1.38            & …                & 3.84      & 1.10            & …                & 1.00             \\ \hline
UGC00634         & NAN            & …              & …               & 5.83           & 1.57            & …                & 2.43      & 0.49            & …                & 2.00             \\ \hline
UGC00731         & 0.22           & 0.50            & 0.00             & 0.08           & 3.79            & …                & 6.42      & 2.39            & …                & 1.00             \\ \hline
UGC00891         & NAN            & …           & …            & 1.45           & 1.34            & …                & 25.16     & 0.32            & …                & 2.00             \\ \hline
UGC01230         & 1.18           & 0.52            & 0.00             & 0.80           & 1.71            & …                & 2.95      & 0.72            & …                & 1.00             \\ \hline
UGC01281         & 0.14           & 0.50            & 0.00             & 0.34           & 1.42            & …                & 0.24      & 0.39            & …                & 1.00             \\ \hline
UGC02023         & NAN            &…            & …             & 0.06           & 0.72            & …                & 1.15      & 0.49            & …                & 2.00             \\ \hline
UGC02259         & 0.96           & 0.50            & 0.00             & 4.66           & 1.81            & …                & 7.22      & 1.14            & …                & 2.00             \\ \hline
UGC02455         & 1.07           & 0.45            & 0.00             & 0.90           & 0.28            & …                & 6.55      & 0.33            & …                & 3.00             \\ \hline
UGC02487         & 4.86           & 0.82            & 0.54             & 4.18           & 1.24            & 0.77             & 4.48      & 1.83            & 0.91             & 1.00             \\ \hline
UGC02885         & 0.84           & 0.47            & 1.00             & 2.14           & 0.50            & 0.98             & 0.86      & 0.45            & 0.97             & 1.00             \\ \hline
UGC02916         & 11.58          & 1.06            & 0.50             & 11.36          & 1.41            & 0.71             & 11.65     & 1.57            & 0.73             & 2.00             \\ \hline
UGC02953         & 4.97           & 0.50            & 0.60             & 5.69           & 0.74            & 0.78             & 5.66      & 0.61            & 0.62             & 2.00             \\ \hline
UGC03205         & 3.82           & 0.67            & 0.88             & 2.99           & 0.68            & 1.07             & 4.20      & 0.73            & 1.32             & 1.00             \\ \hline
UGC03546         & 0.99           & 0.61            & 0.53             & 1.14           & 0.61            & 0.60             & 0.91      & 0.68            & 0.51             & 1.00             \\ \hline
UGC03580         & 2.40           & 0.94            & 0.32             & 2.20           & 0.48            & 0.41             & 2.29      & 0.29            & 0.11             & 2.00             \\ \hline
UGC04278         & 0.58           & 0.66            & 0.00             & 0.92           & 1.00            & …                & 2.60      & 0.53            & …                & 1.00             \\ \hline
UGC04305         & 1.45           & 0.54            & 0.00             & NAN            & …               & …                & 2.02      & 0.71            & …                & 3.00             \\ \hline
UGC04325         & 2.89           & 0.52            & 0.00             & 3.72           & 1.87            & …                & 9.43      & 0.94            & …                & 1.00             \\ \hline
UGC04483         & 0.54           & 0.50            & 0.00             & 0.52           & 1.28            & …                & 0.87      & 0.43            & …                & 2.00             \\ \hline
UGC04499         & 0.46           & 0.46            & 0.00             & 1.54           & 1.14            & …                & 1.78      & 0.51            & …                & 1.00             \\ \hline
UGC05005         & 0.05           & 0.50            & 0.00             & 0.09           & 1.03            & …                & 0.32      & 0.45            & …                & 1.00             \\ \hline
UGC05253         & 3.79           & 0.55            & 0.72             & 5.65           & 0.53            & 0.72             & 4.75      & 0.63            & 0.69             & 2.00             \\ \hline
UGC05414         & 0.37           & 0.52            & 0.00             & 0.19           & 1.06            & …                & 1.30      & 0.41            & …                & 1.00             \\ \hline
UGC05716         & 3.07           & 0.50            & 0.00             & 3.81           & 1.52            & …                & 5.66      & 1.41            & …                & 2.00             \\ \hline
UGC05721         & 0.94           & 0.50            & 0.00             & 0.84           & 1.11            & …                & 1.82      & 0.62            & …                & 1.00             \\ \hline
UGC05750         & 0.26           & 0.49            & 0.00             & 0.47           & 1.57            & …                & 1.35      & 0.48            & …                & 1.00             \\ \hline
UGC05764         & 7.05           & 0.49            & 0.00             & 9.45           & 3.70            & …                & 16.18     & 3.83            & …                & 2.00             \\ \hline
UGC05829         & 0.19           & 0.51            & 0.00             & 0.07           & 1.85            & …                & 0.45      & 0.60            & …                & 2.00             \\ \hline
UGC05918         & 0.03           & 0.50            & 0.00             & 0.21           & 2.31            & …                & 0.94      & 0.54            & …                & 2.00             \\ \hline
UGC05986         & 1.72           & 0.56            & 0.00             & 1.69           & 1.15            & …                & 4.00      & 0.31            & …                & 2.00             \\ \hline
UGC05999         & NAN            &…           & …            & 3.56           & 1.27            & …                & 5.69      & 0.48            & …                & 2.00             \\ \hline
UGC06399         & 0.15           & 0.51            & 0.00             & 0.22           & 1.43            & …                & 0.52      & 0.53            & …                & 1.00             \\ \hline
UGC06446         & 0.22           & 0.50            & 0.00             & 0.17           & 1.69            & …                & 1.00      & 1.04            & …                & 1.00             \\ \hline
UGC06614         & 0.18           & 0.50            & 0.71             & 1.20           & 0.72            & 0.59             & 1.16      & 0.51            & 0.50             & 1.00             \\ \hline
UGC06628         & 0.31           & 0.50            & 0.00             & NAN            & …               & …                & 0.85      & 0.52            & …                & 2.00             \\ \hline
UGC06667         & 0.39           & 0.51            & 0.00             & 2.41           & 3.77            & …                & 5.36      & 1.00            & …                & 1.00             \\ \hline
UGC06786         & 0.69           & 0.35            & 0.78             & 0.69           & 0.50            & 0.66             & 1.39      & 0.27            & 0.34             & 1.00             \\ \hline
UGC06787         & 18.09          & 0.90            & 0.19             & 24.78          & 0.73            & 0.57             & 20.81     & 0.45            & 0.28             & 2.00             \\ \hline
UGC06818         & 2.10           & 0.45            & 0.00             & 1.24           & 0.53            & …                & 5.39      & 0.29            & …                & 2.00             \\ \hline
UGC06917         & 0.39           & 0.48            & 0.00             & 1.09           & 1.12            & …                & 1.32      & 0.54            & …                & 1.00             \\ \hline
UGC06923         & 1.97           & 0.48            & 0.00             & 0.88           & 0.80            & …                & 1.62      & 0.42            & …                & 2.00             \\ \hline
UGC06930         & 0.22           & 0.52            & 0.00             & 0.55           & 1.25            & …                & 1.23      & 0.63            & …                & 1.00             \\ \hline
UGC06973         & 1.45           & 0.31            & 0.60             & 0.41           & 0.36            & 0.81             & 15.58     & 0.17            & 0.39             & 3.00             \\ \hline
UGC06983         & 0.58           & 0.51            & 0.00             & 0.77           & 1.26            & …                & 1.39      & 0.77            & …                & 1.00             \\ \hline
UGC07089         & 0.23           & 0.51            & 0.00             & 0.15           & 0.95            & …                & 0.43      & 0.36            & …                & 2.00             \\ \hline
UGC07125         & 0.40           & 0.50            & 0.00             & 1.18           & 1.08            & …                & 1.60      & 0.92            & …                & 1.00             \\ \hline
UGC07151         & 1.58           & 0.56            & 0.00             & 1.30           & 1.15            & …                & 3.75      & 0.50            & …                & 1.00             \\ \hline
UGC07232         & NAN            & …            & …             & 0.76           & 0.82            & …                & 6.17      & 0.46            & …                & 2.00             \\ \hline
UGC07261         & 0.16           & 0.47            & 0.00             & 1.46           & 1.22            & …                & 0.83      & 0.56            & …                & 2.00             \\ \hline
UGC07323         & 0.43           & 0.51            & 0.00             & 0.26           & 0.97            & …                & 0.66      & 0.41            & …                & 1.00             \\ \hline
UGC07399         & 0.36           & 0.50            & 0.00             & 1.00           & 1.54            & …                & 1.90      & 0.59            & …                & 1.00             \\ \hline
UGC07524         & 0.34           & 0.53            & 0.00             & 1.48           & 1.59            & …                & 1.84      & 0.79            & …                & 1.00             \\ \hline
UGC07559         & 0.26           & 0.50            & 0.00             & 0.21           & 0.97            & …                & 2.60      & 0.31            & …                & 2.00             \\ \hline
UGC07577         & 0.22           & 0.49            & 0.00             & 0.06           & 0.67            & …                & 5.79      & 0.24            & …                & 2.00             \\ \hline
UGC07603         & 0.47           & 0.48            & 0.00             & 0.52           & 1.07            & …                & 1.77      & 0.34            & …                & 1.00             \\ \hline
UGC07608         & 0.14           & 0.50            & 0.00             & 0.30           & 2.02            & …                & 0.73      & 0.48            & …                & 1.00             \\ \hline
UGC07690         & 0.87           & 0.53            & 0.00             & 0.40           & 1.05            & …                & 1.53      & 0.60            & …                & 2.00             \\ \hline
UGC07866         & 0.12           & 0.50            & 0.00             & 0.07           & 1.26            & …                & 0.26      & 0.45            & …                & 2.00             \\ \hline
UGC08286         & 1.00           & 0.52            & 0.00             & 2.66           & 1.66            & …                & 2.64      & 1.05            & …                & 1.00             \\ \hline
UGC08490         & 0.21           & 0.51            & 0.00             & 0.15           & 1.26            & …                & 0.34      & 0.86            & …                & 1.00             \\ \hline
UGC08550         & 0.55           & 0.50            & 0.00             & 0.70           & 1.35            & …                & 1.55      & 0.74            & …                & 1.00             \\ \hline
UGC08699         & 0.87           & 0.75            & 0.59             & 0.95           & 0.55            & 0.76             & 0.99      & 0.63            & 0.70             & 2.00             \\ \hline
UGC08837         & 0.68           & 0.49            & 0.00             & 0.69           & 0.78            & …                & 2.35      & 0.20            & …                & 2.00             \\ \hline
UGC09037         & 1.09           & 0.30            & 0.00             & 1.55           & 0.58            & …                & 2.26      & 0.20            & …                & 2.00             \\ \hline
UGC09133         & 6.78           & 0.65            & 0.53             & 7.50           & 0.77            & 0.72             & 6.94      & 1.64            & 1.10             & 1.00             \\ \hline
UGC09992         & NAN            & …            & …            & 0.04           & 1.30            & …                & 1.08      & 0.51            & …                & 2.00             \\ \hline
UGC10310         & 0.64           & 0.51            & 0.00             & 0.12           & 1.54            & …                & 1.76      & 0.62            & …                & 1.00             \\ \hline
UGC11455         & 2.40           & 0.41            & 0.00             & 3.98           & 0.76            & …                & 6.55      & 0.38            & …                & 1.00             \\ \hline
UGC11557         & 0.67           & 0.48            & 0.00             & 0.91           & 0.60            & …                & 3.18      & 0.42            & …                & 2.00             \\ \hline
UGC11820         & 3.46           & 1.19            & 0.00             & 1.44           & 1.24            & …                & 1.99      & 1.01            & …                & 1.00             \\ \hline
UGC11914         & 0.65           & 0.42            & 0.75             & 1.61           & 0.07            & 0.89             & 1.73      & 0.22            & 0.48             & 1.00             \\ \hline
UGC12506         & 0.29           & 0.59            & 0.00             & 1.18           & 1.35            & …                & 1.98      & 1.12            & …                & 2.00             \\ \hline
UGC12632         & 0.17           & 0.50            & 0.00             & 0.21           & 1.86            & …                & 1.80      & 1.08            & …                & 1.00             \\ \hline
UGC12732         & 0.24           & 0.48            & 0.00             & 0.18           & 1.51            & …                & 0.50      & 1.07            & …                & 1.00             \\ \hline
UGCA281          & 0.38           & 0.49            & 0.00             & 0.28           & 1.23            & …                & 0.47      & 0.37            & …                & 3.00             \\ \hline
UGCA442          & 1.60           & 0.50            & 0.00             & 0.86           & 2.28            & …                & 7.65      & 0.44            & …                & 1.00             \\ \hline
UGCA444          & 0.22           & 0.50            & 0.00             & 0.10           & 3.78            & …                & 0.33      & 0.49            & …                & 2.00             \\ \hline
 \end{longtable}
\end{center}
\end{landscape}

 \clearpage
\twocolumn


\section*{Data Availability}
The code used to generate all figures and fit results is publicly available at \url{https://github.com/Cisneros-Galaxy/RCFM}. The rotation curve data and photometric models used in this paper are from the SPARC database available at \url{http://astroweb.cwru.edu/SPARC/}. The few other datasets used as examples are available at the cited reference.



\bibliographystyle{mnras}
\bibliography{LCM} 

\begin{thebibliography}{}
\makeatletter
\relax
\def\mn@urlcharsother{\let\do\@makeother \do\$\do\&\do\#\do\^\do\_\do\%\do\~}
\def\mn@doi{\begingroup\mn@urlcharsother \@ifnextchar [ {\mn@doi@} {\mn@doi@[]}}
\def\mn@doi@[#1]#2{\def\@tempa{#1}\ifx\@tempa\@empty \href {http://dx.doi.org/#2} {doi:#2}\else \href {http://dx.doi.org/#2} {#1}\fi \endgroup}
\def\mn@eprint#1#2{\mn@eprint@#1:#2::\@nil}
\def\mn@eprint@arXiv#1{\href {http://arxiv.org/abs/#1} {{\tt arXiv:#1}}}
\def\mn@eprint@dblp#1{\href {http://dblp.uni-trier.de/rec/bibtex/#1.xml} {dblp:#1}}
\def\mn@eprint@#1:#2:#3:#4\@nil{\def\@tempa {#1}\def\@tempb {#2}\def\@tempc {#3}\ifx \@tempc \@empty \let \@tempc \@tempb \let \@tempb \@tempa \fi \ifx \@tempb \@empty \def\@tempb {arXiv}\fi \@ifundefined {mn@eprint@\@tempb}{\@tempb:\@tempc}{\expandafter \expandafter \csname mn@eprint@\@tempb\endcsname \expandafter{\@tempc}}}

\bibitem[\protect\citeauthoryear{Battaglia, Fraternali, Oosterloo,   \& Sancisi}{Battaglia et~al.}{2006}]{Batt}
Battaglia G.,  Fraternali F.,  Oosterloo T.,    Sancisi R.,  2006, A\& A, 447, 49

\bibitem[\protect\citeauthoryear{Bekenstein}{Bekenstein}{2004}]{PhysRevDBekenstein2004}
Bekenstein J.~D.,  2004, \mn@doi [Phys. Rev. D] {10.1103/PhysRevD.70.083509}, 70, 083509

\bibitem[\protect\citeauthoryear{Bell \& de Jong}{Bell \& de~Jong}{2001}]{BelldYong}
Bell E.~F.,  de Jong R.~S.,  2001, \mn@doi [ApJ] {10.1086/319728}, \href {http://adsabs.harvard.edu/abs/2001ApJ...550..212B} {550, 212}

\bibitem[\protect\citeauthoryear{Bosma}{Bosma}{1981}]{Bosma}
Bosma A.,  1981, AJ, 86, 1791

\bibitem[\protect\citeauthoryear{{Boyer}}{{Boyer}}{2011}]{2011AmJPh..79..644B}
{Boyer} T.~H.,  2011, \mn@doi [American Journal of Physics] {10.1119/1.3534842}, \href {https://ui.adsabs.harvard.edu/abs/2011AmJPh..79..644B} {79, 644}

\bibitem[\protect\citeauthoryear{{Casertano}}{{Casertano}}{1983}]{1983MNRAS.203..735C}
{Casertano} S.,  1983, \mn@doi [\mnras] {10.1093/mnras/203.3.735}, \href {https://ui.adsabs.harvard.edu/abs/1983MNRAS.203..735C} {203, 735}

\bibitem[\protect\citeauthoryear{Cebri\'an}{Cebri\'an}{2022}]{Cebrian:2022brv}
Cebri\'an S.,  2022, in {10th Symposium on Large TPCs for Low-Energy Rare Event Detection}.  (\mn@eprint {arXiv} {2205.06833})

\bibitem[\protect\citeauthoryear{Chatterjee}{Chatterjee}{1987}]{Chatterjee}
Chatterjee T.,  1987, Ap\&SS, 139, 243

\bibitem[\protect\citeauthoryear{Conroy, Gunn  \& White}{Conroy et~al.}{2009}]{Conroy}
Conroy C.,  Gunn J.,   White M.,  2009, ApJ, 699, 486

\bibitem[\protect\citeauthoryear{Cooperstock \& Tieu}{Cooperstock \& Tieu}{2007}]{doi:10.1142/S0217751X0703666X}
Cooperstock F.~I.,  Tieu S.,  2007, \mn@doi [International Journal of Modern Physics A] {10.1142/S0217751X0703666X}, 22, 2293

\bibitem[\protect\citeauthoryear{Desmond, Bartlett  \& Ferreira}{Desmond et~al.}{2023}]{10.1093/mnras/stad597}
Desmond H.,  Bartlett D.~J.,   Ferreira P.~G.,  2023, \mn@doi [Monthly Notices of the Royal Astronomical Society] {10.1093/mnras/stad597}, 521, 1817

\bibitem[\protect\citeauthoryear{Dutton, Courteau, de Jong  \& Carignan}{Dutton et~al.}{2005}]{Dutton_2005}
Dutton A.~A.,  Courteau S.,  de Jong R.,   Carignan C.,  2005, \mn@doi [The Astrophysical Journal] {10.1086/426375}, 619, 218

\bibitem[\protect\citeauthoryear{Dutton, Macciò, Obreja  \& Buck}{Dutton et~al.}{2019}]{10.1093/mnras/stz531}
Dutton A.~A.,  Macciò A.~V.,  Obreja A.,   Buck T.,  2019, \mn@doi [Monthly Notices of the Royal Astronomical Society] {10.1093/mnras/stz531}, 485, 1886

\bibitem[\protect\citeauthoryear{Famaey \& McGaugh}{Famaey \& McGaugh}{2012}]{Famaey2012}
Famaey B.,  McGaugh S.~S.,  2012, \mn@doi [Living Reviews in Relativity] {10.12942/lrr-2012-10}, 15, 10

\bibitem[\protect\citeauthoryear{{Fich} \& {Tremaine}}{{Fich} \& {Tremaine}}{1991}]{1991ARA&A..29..409F}
{Fich} M.,  {Tremaine} S.,  1991, \mn@doi [\araa] {10.1146/annurev.aa.29.090191.002205}, \href {https://ui.adsabs.harvard.edu/abs/1991ARA&A..29..409F} {29, 409}

\bibitem[\protect\citeauthoryear{Fraternali, Sancisi,   \& Kamphuis}{Fraternali et~al.}{2011}]{Frat1}
Fraternali F.,  Sancisi R.,    Kamphuis P.,  2011, A\&A, http://arxiv.org/abs/1105.3867

\bibitem[\protect\citeauthoryear{{Freeman}}{{Freeman}}{1970}]{Freeman}
{Freeman} K.~C.,  1970, \mn@doi [\apj] {10.1086/150474}, \href {https://ui.adsabs.harvard.edu/abs/1970ApJ...160..811F} {160, 811}

\bibitem[\protect\citeauthoryear{Gentile, Farnaey  \& de Blok}{Gentile et~al.}{2011}]{Gent}
Gentile G.,  Farnaey B.,   de Blok W.,  2011, A\&A, 527, A76

\bibitem[\protect\citeauthoryear{Gentile et~al.,}{Gentile et~al.}{2013}]{Maria1}
Gentile G.,  et~al., 2013, \mn@doi [A\&A] {10.1051/0004-6361/201321116}, \href {http://adsabs.harvard.edu/abs/2013A26A...554A.125G} {554, A125}

\bibitem[\protect\citeauthoryear{{Hur{\'e}} \& {Hersant}}{{Hur{\'e}} \& {Hersant}}{2011}]{2011A&A...531A..36H}
{Hur{\'e}} J.~M.,  {Hersant} F.,  2011, \mn@doi [\aap] {10.1051/0004-6361/201015854}, \href {https://ui.adsabs.harvard.edu/abs/2011A&A...531A..36H} {531, A36}

\bibitem[\protect\citeauthoryear{Jackson}{Jackson}{1999}]{Jack}
Jackson J.,  1999, Classical Electrodynamics, 3rd edn.
John Wiley \& Sons, Inc., New Jersey, USA

\bibitem[\protect\citeauthoryear{{Jetzer}}{{Jetzer}}{2017}]{jetzer}
{Jetzer} P.,  2017, Physik-Institut der Universitat Zurich

\bibitem[\protect\citeauthoryear{{Jiao, Yongjun}, {Hammer, Fran\c{c}ois}, {Wang, Haifeng}, {Wang, Jianling}, {Amram, Philippe}, {Chemin, Laurent}  \& {Yang, Yanbin}}{{Jiao, Yongjun} et~al.}{2023}]{jiao2023detection}
{Jiao, Yongjun} {Hammer, Fran\c{c}ois} {Wang, Haifeng} {Wang, Jianling} {Amram, Philippe} {Chemin, Laurent}  {Yang, Yanbin} 2023, \mn@doi [A\&A] {10.1051/0004-6361/202347513}, 678, A208

\bibitem[\protect\citeauthoryear{Keller \& Wadsley}{Keller \& Wadsley}{2017}]{Keller_2017}
Keller B.~W.,  Wadsley J.~W.,  2017, \mn@doi [The Astrophysical Journal Letters] {10.3847/2041-8213/835/1/l17}, 835, L17

\bibitem[\protect\citeauthoryear{{Lelli}, {McGaugh}  \& {Schombert}}{{Lelli} et~al.}{2016}]{2016Lelli}
{Lelli} F.,  {McGaugh} S.~S.,   {Schombert} J.~M.,  2016, \mn@doi [AJ] {10.3847/0004-6256/152/6/157}, \href {http://adsabs.harvard.edu/abs/2016AJ....152..157L} {152, 157}

\bibitem[\protect\citeauthoryear{Lelli, McGaugh, Schombert  \& Pawlowski}{Lelli et~al.}{2017}]{Lelli_2017}
Lelli F.,  McGaugh S.~S.,  Schombert J.~M.,   Pawlowski M.~S.,  2017, \mn@doi [The Astrophysical Journal] {10.3847/1538-4357/836/2/152}, 836, 152

\bibitem[\protect\citeauthoryear{Li, Lelli, McGaugh  \& Schombert}{Li et~al.}{2018}]{Li_2018}
Li P.,  Lelli F.,  McGaugh S.,   Schombert J.,  2018, \mn@doi [Astronomy I\& Astrophysics] {10.1051/0004-6361/201732547}, 615, A3

\bibitem[\protect\citeauthoryear{Maschberger, Bonnell, Clarke  \& Moraux}{Maschberger et~al.}{2014}]{10.1093/mnras/stt2403}
Maschberger T.,  Bonnell I.~A.,  Clarke C.~J.,   Moraux E.,  2014, \mn@doi [Monthly Notices of the Royal Astronomical Society] {10.1093/mnras/stt2403}, 439, 234

\bibitem[\protect\citeauthoryear{McGaugh}{McGaugh}{1999}]{1999McGaugh}
McGaugh S.,  1999, Conference Proceedings, 182

\bibitem[\protect\citeauthoryear{{McGaugh}}{{McGaugh}}{2004}]{2004ApJ...609..652M}
{McGaugh} S.~S.,  2004, \mn@doi [\apj] {10.1086/421338}, \href {https://ui.adsabs.harvard.edu/abs/2004ApJ...609..652M} {609, 652}

\bibitem[\protect\citeauthoryear{McGaugh}{McGaugh}{2008}]{McGaugh_2008}
McGaugh S.~S.,  2008, \mn@doi [The Astrophysical Journal] {10.1086/589148}, 683, 137

\bibitem[\protect\citeauthoryear{McGaugh}{McGaugh}{2014}]{McGaugh_2014}
McGaugh S.,  2014, \mn@doi [Galaxies] {10.3390/galaxies2040601}, 2, 601–622

\bibitem[\protect\citeauthoryear{McGaugh}{McGaugh}{2021}]{MCGAUGH2021220}
McGaugh S.~S.,  2021, \mn@doi [Studies in History and Philosophy of Science] {https://doi.org/10.1016/j.shpsa.2021.05.008}, 88, 220

\bibitem[\protect\citeauthoryear{McGaugh, Lelli  \& Schombert}{McGaugh et~al.}{2016}]{McGaugh2016RAR}
McGaugh S.~S.,  Lelli F.,   Schombert J.~M.,  2016, \mn@doi [Phys. Rev. Lett.] {10.1103/PhysRevLett.117.201101}, 117, 201101

\bibitem[\protect\citeauthoryear{{Milgrom}}{{Milgrom}}{1983}]{Milgrom}
{Milgrom} M.,  1983, \mn@doi [ApJ] {10.1086/161131}, \href {http://adsabs.harvard.edu/abs/1983ApJ...270..371M} {270, 371}

\bibitem[\protect\citeauthoryear{{Misner}, {Thorne}  \& {Wheeler}}{{Misner} et~al.}{1973}]{MTW}
{Misner} C.~W.,  {Thorne} K.~S.,   {Wheeler} J.~A.,  1973, Gravitation

\bibitem[\protect\citeauthoryear{{Mistele}, {McGaugh}  \& {Hossenfelder}}{{Mistele} et~al.}{2022}]{2022A&A...664A..40M}
{Mistele} T.,  {McGaugh} S.,   {Hossenfelder} S.,  2022, \mn@doi [\aap] {10.1051/0004-6361/202243216}, \href {https://ui.adsabs.harvard.edu/abs/2022A&A...664A..40M} {664, A40}

\bibitem[\protect\citeauthoryear{Naidu et~al.,}{Naidu et~al.}{2022}]{Naidu_2022}
Naidu R.~P.,  et~al., 2022, \mn@doi [The Astrophysical Journal Letters] {10.3847/2041-8213/ac9b22}, 940, L14

\bibitem[\protect\citeauthoryear{{Navarro}, {Frenk}  \& {White}}{{Navarro} et~al.}{1996}]{1996ApJ...462..563N}
{Navarro} J.~F.,  {Frenk} C.~S.,   {White} S. D.~M.,  1996, \mn@doi [\apj] {10.1086/177173}, \href {https://ui.adsabs.harvard.edu/abs/1996ApJ...462..563N} {462, 563}

\bibitem[\protect\citeauthoryear{{Navarro}, {Ben{\'\i}tez-Llambay}, {Fattahi}, {Frenk}, {Ludlow}, {Oman}, {Schaller}  \& {Theuns}}{{Navarro} et~al.}{2017}]{2017MNRAS.471.1841N}
{Navarro} J.~F.,  {Ben{\'\i}tez-Llambay} A.,  {Fattahi} A.,  {Frenk} C.~S.,  {Ludlow} A.~D.,  {Oman} K.~A.,  {Schaller} M.,   {Theuns} T.,  2017, \mn@doi [\mnras] {10.1093/mnras/stx170510.48550/arXiv.1612.06329}, \href {https://ui.adsabs.harvard.edu/abs/2017MNRAS.471.1841N} {471, 1841}

\bibitem[\protect\citeauthoryear{Oh, De~Blok, Walter, Brinks  \& Kennicutt}{Oh et~al.}{2008}]{oh2008high}
Oh S.-H.,  De~Blok W.,  Walter F.,  Brinks E.,   Kennicutt R.~C.,  2008, The Astronomical Journal, 136, 2761

\bibitem[\protect\citeauthoryear{{Persic}, {Salucci}  \& {Stel}}{{Persic} et~al.}{1996}]{salucci}
{Persic} M.,  {Salucci} P.,   {Stel} F.,  1996, MNRAS, \href {http://adsabs.harvard.edu/abs/1996MNRAS.281...27P} {281, 27}

\bibitem[\protect\citeauthoryear{Pomar\`ede, Tully, Graziani, Courtois, Hoffman  \& Lezmy}{Pomar\`ede et~al.}{2020}]{Pomarede:2020pme}
Pomar\`ede D.,  Tully R.~B.,  Graziani R.,  Courtois H.~M.,  Hoffman Y.,   Lezmy J.,  2020, \mn@doi [Astrophys. J.] {10.3847/1538-4357/ab9952}, 897, 133

\bibitem[\protect\citeauthoryear{Rindler}{Rindler}{2013}]{rindler2013essential}
Rindler W.,  2013, Essential Relativity: Special, General, and Cosmological.
Springer New York, \url {https://books.google.com/books?id=WTfnBwAAQBAJ}

\bibitem[\protect\citeauthoryear{Rubin, Thonnard  \& Ford}{Rubin et~al.}{1978}]{1978Rubin}
Rubin V.~C.,  Thonnard N.,   Ford Jr. W.~K.,  1978, \mn@doi [ApJ] {10.1086/182804}, \href {http://adsabs.harvard.edu/abs/1978ApJ...225L.107R} {225, L107}

\bibitem[\protect\citeauthoryear{Rubin, Ford  \& Thonnard}{Rubin et~al.}{1980}]{Rub}
Rubin V.,  Ford W.,   Thonnard N.,  1980, ApJ, 238, 471

\bibitem[\protect\citeauthoryear{Salucci, Lapi, Tonini, Gentile, Yegorova  \& Klein}{Salucci et~al.}{2007}]{10.1111/j.1365-2966.2007.11696.x}
Salucci P.,  Lapi A.,  Tonini C.,  Gentile G.,  Yegorova I.,   Klein U.,  2007, \mn@doi [Monthly Notices of the Royal Astronomical Society] {10.1111/j.1365-2966.2007.11696.x}, 378, 41

\bibitem[\protect\citeauthoryear{{Sanders}}{{Sanders}}{2010}]{2010dmp..book.....S}
{Sanders} R.~H.,  2010, {The Dark Matter Problem: A Historical Perspective}

\bibitem[\protect\citeauthoryear{Schombert, McGaugh  \& Lelli}{Schombert et~al.}{2018}]{10.1093/mnras/sty3223}
Schombert J.,  McGaugh S.,   Lelli F.,  2018, \mn@doi [Monthly Notices of the Royal Astronomical Society] {10.1093/mnras/sty3223}, 483, 1496

\bibitem[\protect\citeauthoryear{{Schwarzschild}}{{Schwarzschild}}{1954}]{1954AJ.....59..273S}
{Schwarzschild} M.,  1954, \mn@doi [\aj] {10.1086/107013}, \href {https://ui.adsabs.harvard.edu/abs/1954AJ.....59..273S} {59, 273}

\bibitem[\protect\citeauthoryear{{Sofue}}{{Sofue}}{2013}]{Sofue}
{Sofue} Y.,  2013, {Mass Distribution and Rotation Curve in the Galaxy}.
Oswalt, T.~D. and Gilmore, G., p.~985, \mn@doi{10.1007/978-94-007-5612-019}

\bibitem[\protect\citeauthoryear{Tecchiolli}{Tecchiolli}{2019}]{universe5100206}
Tecchiolli M.,  2019, \mn@doi [Universe] {10.3390/universe5100206}, 5

\bibitem[\protect\citeauthoryear{Tully, Courtois, Hoffman  \& Pomar\`ede}{Tully et~al.}{2014}]{Tully:2014gfa}
Tully R.~B.,  Courtois H.,  Hoffman Y.,   Pomar\`ede D.,  2014, \mn@doi [Nature] {10.1038/nature13674}, 513, 71

\bibitem[\protect\citeauthoryear{Wald}{Wald}{1984}]{Wald}
Wald R.,  1984, General Relativity.
University of Chicago Press, Chicago, IL, USA

\bibitem[\protect\citeauthoryear{Xue et~al.}{Xue et~al.}{2008}]{Xue}
Xue X.,  et~al., 2008, \mn@doi [AJ.] {10.1086/589500}, 684, 1143

\bibitem[\protect\citeauthoryear{Zel'dovich}{Zel'dovich}{1968}]{YaBZel'dovich_1968}
Zel'dovich Y.~B.,  1968, \mn@doi [Soviet Physics Uspekhi] {10.1070/PU1968v011n03ABEH003927}, 11, 381

\bibitem[\protect\citeauthoryear{de Blok \& McGaugh}{de~Blok \& McGaugh}{1997}]{deBlok}
de Blok W.,  McGaugh S.,  1997, MNRAS, 290, 533

\bibitem[\protect\citeauthoryear{de Blok, Walter,   \& Brinks}{de~Blok et~al.}{2008a}]{Blok}
de Blok W.,  Walter F.,    Brinks E.,  2008a, AJ, 136, 2648

\bibitem[\protect\citeauthoryear{de Blok, Walter,   \& Brinks}{de~Blok et~al.}{2008b}]{Blok1}
de Blok W.,  Walter F.,    Brinks E.,  2008b, AJ, 136, 2648

\bibitem[\protect\citeauthoryear{{de Vaucouleurs}}{{de Vaucouleurs}}{1959}]{1959HDP....53..275D}
{de Vaucouleurs} G.,  1959, \mn@doi [Handbuch der Physik] {10.1007/978-3-642-45932-0_7}, \href {https://ui.adsabs.harvard.edu/abs/1959HDP....53..275D} {53, 275}

\bibitem[\protect\citeauthoryear{{van Albada}, {Bahcall}, {Begeman}  \& {Sancisi}}{{van Albada} et~al.}{1985}]{1985ApJAlbada}
{van Albada} T.~S.,  {Bahcall} J.~N.,  {Begeman} K.,   {Sancisi} R.,  1985, \mn@doi [\apj] {10.1086/163375}, \href {http://adsabs.harvard.edu/abs/1985ApJ...295..305V} {295, 305}

\bibitem[\protect\citeauthoryear{Željko Ivezić et~al.,}{Željko Ivezić et~al.}{2019}]{Ivezić_2019}
Željko Ivezić et~al., 2019, \mn@doi [The Astrophysical Journal] {10.3847/1538-4357/ab042c}, 873, 111

\makeatother
\end{thebibliography}





\bsp	
\label{lastpage}
\end{document}